\documentclass[%
superscriptaddress,
 aps,
 reprint
]{revtex4-2}

\usepackage{graphicx}% Include figure files
\usepackage{dcolumn}% Align table columns on decimal point
\usepackage{bm}% bold math
\usepackage{tablefootnote}
\usepackage{array,booktabs,siunitx}
\usepackage{tabularx}
\usepackage{xcolor}
\usepackage[utf8]{inputenc}
\usepackage[T1]{fontenc}
\usepackage{mathptmx}
\usepackage{float}
\usepackage{sidecap}
\usepackage{siunitx}
\usepackage{gensymb}
\usepackage{hyperref}

\usepackage{textcomp}%
\usepackage{booktabs}%
\usepackage{listings}%

\begin{document}

\title[]{Infrared Spectroscopy for Diagnosing Superlattice Minibands in Magic-angle Twisted Bilayer Graphene}

\author{Geng Li}
\affiliation{ICFO-Institut de Ci\`{e}ncies Fot\`{o}niques, The Barcelona Institute of Science and Technology, Av. Carl Friedrich Gauss 3, Castelldefels (Barcelona), 08860, Spain}

\author{Roshan Krishna Kumar}
\email{roshan.krishnakumar@icfo.eu}
\affiliation{ICFO-Institut de Ci\`{e}ncies Fot\`{o}niques, The Barcelona Institute of Science and Technology, Av. Carl Friedrich Gauss 3, Castelldefels (Barcelona), 08860, Spain}

\author{Petr Stepanov}
\affiliation{ICFO-Institut de Ci\`{e}ncies Fot\`{o}niques, The Barcelona Institute of Science and Technology, Av. Carl Friedrich Gauss 3, Castelldefels (Barcelona), 08860, Spain}
\affiliation{Department of Physics and Astronomy, University of Notre Dame, Notre Dame, Indiana, 46556, USA}

\author{Pierre A. Pantaleón}
\author{Zhen Zhan}
\affiliation{Imdea Nanoscience, Faraday 9, Madrid, 28047, Spain}

\author{Hitesh Agarwal}
\affiliation{ICFO-Institut de Ci\`{e}ncies Fot\`{o}niques, The Barcelona Institute of Science and Technology, Av. Carl Friedrich Gauss 3, Castelldefels (Barcelona), 08860, Spain}

\author{Adrien Bercher}
\affiliation{Department of Quantum Matter Physics, University of Geneva, 24 Quai Ernest-Ansermet, Geneva, 1205, Switzerland}

\author{Julien Barrier}
\affiliation{ICFO-Institut de Ci\`{e}ncies Fot\`{o}niques, The Barcelona Institute of Science and Technology, Av. Carl Friedrich Gauss 3, Castelldefels (Barcelona), 08860, Spain}

\author{Kenji Watanabe}
\affiliation{Research Center for Electronic and Optical Materials, National Institute for Materials Science, 1-1 Namiki, Tsukuba, 305-0044, Japan}

\author{Takashi Taniguchi}
\affiliation{Research Center for Materials Nanoarchitectonics, National Institute for Materials Science, 1-1 Namiki, Tsukuba, 305-0044, Japan}

\author{Alexey B. Kuzmenko}
\affiliation{Department of Quantum Matter Physics, University of Geneva, 24 Quai Ernest-Ansermet, Geneva, 1205, Switzerland}

\author{Francisco Guinea}
\affiliation{Imdea Nanoscience, Faraday 9, Madrid, 28047, Spain}
\affiliation{Donostia International Physics Center, Paseo Manuel de Lardizabal 4, San Sebastian, 20018, Spain}

\author{Iacopo Torre}
\affiliation{ICFO-Institut de Ci\`{e}ncies Fot\`{o}niques, The Barcelona Institute of Science and Technology, Av. Carl Friedrich Gauss 3, Castelldefels (Barcelona), 08860, Spain}

\author{Frank H.L. Koppens}
\email{frank.koppens@icfo.eu}
\affiliation{ICFO-Institut de Ci\`{e}ncies Fot\`{o}niques, The Barcelona Institute of Science and Technology, Av. Carl Friedrich Gauss 3, Castelldefels (Barcelona), 08860, Spain}
\affiliation{ICREA-Instituci\'{o} Catalana de Recerca i Estudis Avan\c{c}ats, Barcelona, 08010, Spain}

\begin{abstract}

Twisted bilayer graphene (TBG) represents a highly tunable, strongly correlated electron system owed to its unique flat electronic bands. However, understanding the single-particle band structure alone has been challenging due to complex lattice reconstruction effects and a lack of spectroscopic measurements over a broad energy range. Here, we probe the band structure of TBG around the magic angle using infrared spectroscopy. Our measurements reveal spectral features originating from interband transitions whose energies are uniquely defined by the twist angle. By combining with quantum transport, we connect spectral features over a broad energy range (10-700\,meV) spanning several superlattice minibands and track their evolution with twist angle. We compare our data with calculations of the band structures obtained via the continuum model and find good agreement only when considering a variation of interlayer/intralayer tunneling parameters with the twist angle. Our analysis suggests that the magic angle also shifts due to lattice relaxation, and is better defined for a wide angular range of 0.9\degree-1.1\degree. Our work provides spectroscopic insights into TBG’s band structure and offers an optical fingerprint of the magic angle for screening heterostructures before nanofabrication.

\end{abstract}

\maketitle

\section{Introduction}

Twisted bilayer graphene (TBG) represents a paradigm shift in condensed matter physics, as its electronic structure is highly sensitive to minute changes in the arrangements of atoms induced by the relative twisting of crystallographic axes \cite{andrei2020graphene, balents2020superconductivity}. This high sensitivity allows one to tune a 2D electron gas from a non-interacting Fermi-liquid to a strongly correlated electron system at the so-called magic angle (1.1\degree) \cite{cao2018unconventional, cao2018correlated} , which hosts a multitude of exotic quantum phases \cite{polshyn2019large, lu2019superconductors, stepanov2020untying, saito2020independent, serlin2020intrinsic}. Obtaining the single-particle band structure is one of the most important steps to predicting and understanding those phases, but, until now, has been challenging due to the complexity of the system and sensitivity to external perturbations \cite{andrei2020graphene, balents2020superconductivity}. Although various scanning probe experiments \cite{li2010observation, xie2019spectroscopic, choi2019electronic} combined with theoretical models \cite{dos2007graphene, bistritzer2011moire,tarnopolsky2019origin, carr2019exact,bennett2023twisted} have captured key qualitative features of the TBG band structure, a systematic and quantitative study of the evolution in global properties with twist angle is still missing. Optical spectroscopy offers a versatile approach to studying 2D materials via probing interband transitions over a broad energy range \cite{wang2008gate,  kuzmenko2009determination}, and has also recently proven a powerful technique for the study of graphene superlattices \cite{moon2013optical,yu2019gate, havener2014van, kim2016chiral, han2021accurate, yang2022spectroscopy} and twisted transition-metal dichalcogenide heterostructures\cite{alexeev2019resonantly, jin2019observation, seyler2019signatures, tran2019evidence}, whose optical responses are dramatically modified by the underlying moiré potential.
While TBG close to the magic angle hosts abundant optical transitions \cite{novelli2020optical}, the resonant features lie in the mid-infrared and terahertz frequency range\cite{moon2013optical, deng2020strong,ma2022intelligent,hesp2021nano,hesp2021observation,utama2021visualization,lisi2021observation,sunku2020nano}, making optical spectroscopy challenging because of the large optical wavelength relative to the typical TBG device size. 

\begin{figure*}[]
    \centering
    \includegraphics[width=0.9\textwidth]{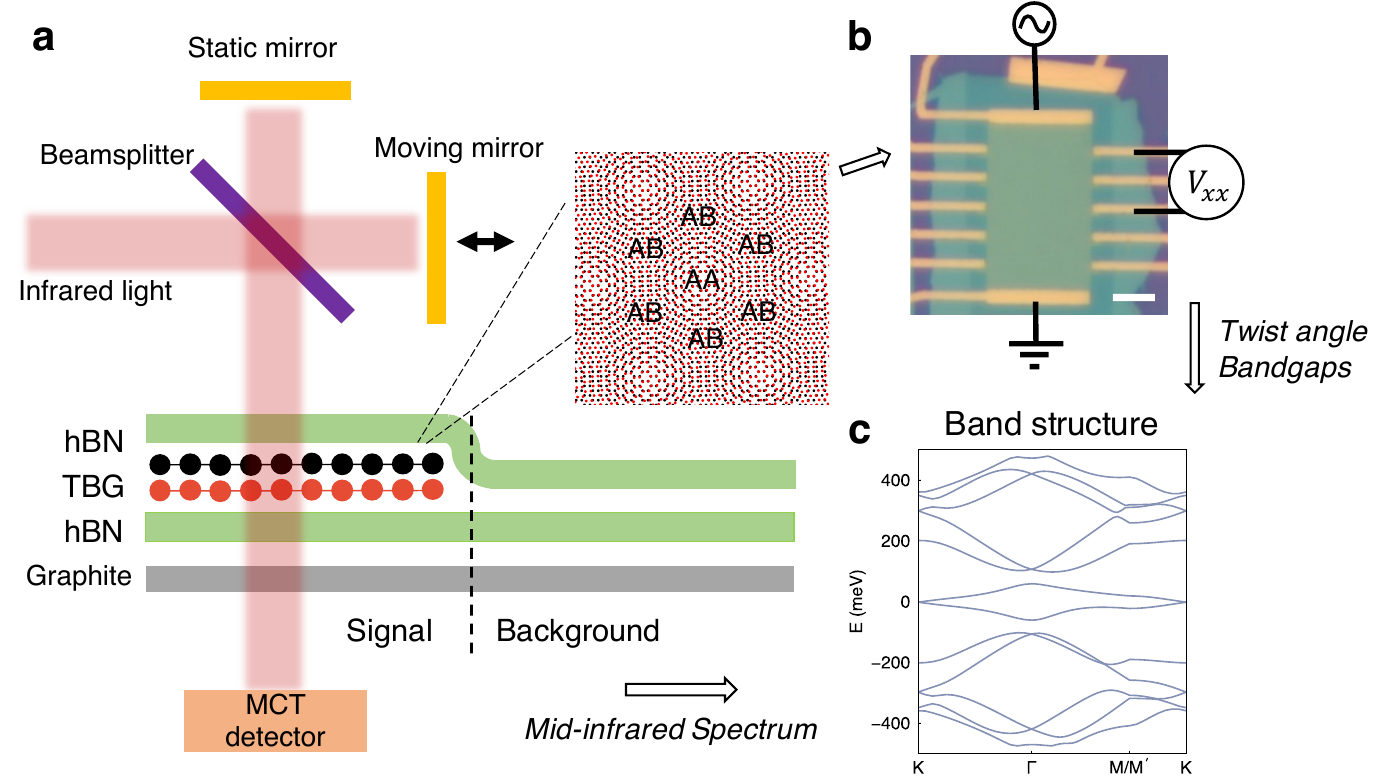}
    \caption{ \textbf{Schematic of work flow. }\textbf{a,} Schematic of the device structure and FTIR measurements. The green, black+red, and grey layers represent hBN, TBG, and graphite layers. The left part of the device consists of hBN/TBG/hBN/graphite and the right part consists of hBN/hBN/graphite. The top and bottom hBN and graphite are kept at exactly the same thicknesses for background normalization. Inset: Moiré superlattice consists of AA and AB regions. \textbf{b,} The signal region of the device is fabricated into a hall bar device after the FTIR measurements. Resistivity is measured by sourcing constant current and sensing voltage drop by four-probe measurements at a base temperature of 10 K. Scale bar, 5 $\mu$m. \textbf{c,} We obtain the band structure of TBG by comparing the calculations of the continuum model to the FTIR and transport measurements.}
    \label{fig:Fig_1}  
    
\end{figure*}

In this study, we combine infrared optical spectroscopy and quantum transport to study clean and large TBG heterostructures with twist angles ranging from 0.94\degree\  to 1.67\degree. During the assembly of our heterostructures, we make reference areas that include the same dielectric materials but exclude the material of interest (Fig. \ref{fig:Fig_1}a). In that way, we can employ a background normalization, which allows us to distinguish infrared spectral features using room temperature Fourier transform infrared spectroscopy (FTIR). Following this, the same heterostructures are fabricated into high-quality quantum transport devices (Fig. \ref{fig:Fig_1}b), allowing us to extract further material parameters, including the twist angle and bandgaps, to relate them to the infrared spectral features. By analyzing the spectral features and quantum transport measurements, we model the single-particle band structure of TBG (Fig. \ref{fig:Fig_1}c) and track its evolution with the twist angle.

%%%%%%%%%%%%%%%%%%%%%%%%%%%%%%%%%%%%%%%%%%
%

We fabricate TBG heterostructures using the “cut-and-stack” method\cite{stepanov2020untying,saito2020independent} (see Methods). We use a few-layer graphite gate because it enables optical spectroscopy measurements thanks to its infrared transparency and provides electrostatic gating in the following transport measurements. The heterostructure (Fig. \ref{fig:Fig_1}a) contains a “signal” region and a “reference” region. The signal region consists of hBN/TBG/hBN/graphite, while the reference region consists of hBN/hBN/graphite, excluding the TBG. Importantly, the layers in both regions are made with the same flakes with uniform thickness. Therefore, we obtain the extinction spectrum of TBG by normalizing the spectrum collected at the signal region with the spectrum from the reference region (1-signal/reference)$\times$100\%.  Both regions are in general around 20$\times$20 \textmu m$^2$, large enough to provide sufficient signal while the signal region maintains a low twist-angle inhomogeneity (Supplementary \ref{sec:angle_inhomogeneity}). The measurements are performed using a commercial FTIR equipped with a microscope and a mercury cadmium telluride (MCT) detector in the transmission mode at room temperature (see Methods for details). 

%%%%%%%%%%%%%%%%%%%%%%%%%%%%%%%%%%%%%%%%%%
%

\section{Infrared spectra of TBG}

To verify that the method for background normalization illustrated in Fig. \ref{fig:Fig_1}a works well, we first perform control experiments of Bernal bilayer graphene whose spectral properties are well known \cite{wang2008gate,zhang2008determination,kuzmenko2009determination}. The grey curve in Fig. \ref{fig:Fig_2}a plots the extinction spectrum, which shows a shoulder-like feature around 400\,meV after which it transitions to a regime of constant universal absorption $\sim$4\%. This behavior is well described by theory\cite{min2009origin,ando2009optical} and corresponds to the onset of interband transitions between parallel bands (Fig. \ref{fig:Fig_2}a inset) unique to Bernal bilayer graphene, resulting in a broadband universal optical absorption. The measured transition energy is also in agreement with previous works that reported values of 350-400 meV\cite{wang2008gate,zhang2008determination,kuzmenko2009determination}. Therefore, we demonstrate the effectiveness of background normalization and its suitability for studying encapsulated graphene heterostructures. Fig. \ref{fig:Fig_2}b plots the resistivity ($\rho_{xx}$) of Bernal bilayer graphene as a function of carrier density, revealing the usual behavior of a peaked response at charge neutrality without additional superlattice features.

We now switch to a TBG device with 1.46\degree\  twist angle. The red curve in Fig. \ref{fig:Fig_2}a plots the extinction spectrum. Strikingly, we observe four distinct peaks labeled 1-4 (indicated by the colored arrows), which we refer to throughout this work, strongly contrasting with the spectrum of Bernal bilayer graphene. Fig. \ref{fig:Fig_2}b (bottom panel) plots the resistivity as a function of carrier density measured in the device fabricated from this heterostructure. In addition to the main peak at charge neutrality, satellite peaks of resistivity are observed at finite doping, where the band edge is reached upon doping 4 electrons/holes per moiré unit cell\cite{cao2018unconventional,cao2018correlated}. 
Temperature-dependent measurements of the high resistance states reveal the expected activated behavior, from which gaps of 50 (46)\,meV for electrons (holes, respectively) are extracted, consistent with previous works\cite{cao2016superlattice,polshyn2019large} (Fig. \ref{fig:Fig_2}b inset). Knowing the carrier density where satellite peaks occur (carrier density determined via Hall effect measurements, shown in supplementary  \ref{sec:angle_extraction}), we determine the twist angle of this heterostructure to be 1.46\degree. The continuum model\cite{morell2010flat,bistritzer2011moire,tarnopolsky2019origin} offers a simple but effective way of describing the band structure of TBG. Using the single-particle band structure calculated from the continuum model (Fig. \ref{fig:Fig_2}d), we calculate the optical conductivity using the Kubo formula\cite{novelli2020optical} (Fig. \ref{fig:Fig_2}c). Quantitatively, we find that the measured resonances shown in Fig. \ref{fig:Fig_2}a align with the calculated resonances depicted in Fig. \ref{fig:Fig_2}c.
%
%%% 

We proceed to study the spectral response of TBG heterostructures with a range of twist angles below and above the magic angle. Fig. \ref{fig:Fig_3}a plots the extinction spectra measured for a range of samples varying from 0.94\degree\ to 1.67\degree\ (twist angles obtained from quantum transport measurements). All the twist angles show distinct spectral features with peaks shifting to lower energies upon lowering the twist angle (traced by dashed lines). This behavior resembles the evolution of the band structure of TBG, in which the interband transition energies shift to lower energy with decreasing twist angle. For twist angles approaching the magic angle, 1.1\degree, we observe a strong increase in band resistivity (Fig. \ref{fig:Fig_2}c) and additional maxima at integer fillings start to manifest, indicative of narrow electronic band consistent with previous works\cite{cao2018correlated,lu2019superconductors,polshyn2019large}. Notably, we find the features at integer fillings persists in our 0.94\degree\  and 0.98\degree\ samples (Supplementary \ref{sec:transport}). %%%% TODO 

Despite such small differences in twist angles between 0.94\degree\  and 0.98\degree\ and between 1.12\degree\ and 1.16\degree, the FTIR measurements show notable energy shifts in the spectral peaks. Such a large sensitivity raises the question as to whether our spectra are skewed by twist-angle inhomogeneity commonly observed in TBG \cite{uri2020mapping, grover2022chern}. However, from measurements of the two-probe resistance between different contact pairs, we extract a twist-angle inhomogeneity within $\pm$0.02° (Supplementary \ref{sec:angle_inhomogeneity}). We attribute this to the high quality of our heterostructures which is evident from the atomic force microscopy image showing no wrinkles and only a few small bubbles (Supplementary \ref{sec:Alignment_hBN}). As an additional check, we measure the extinction spectra of the TBG heterostructures during various stages of the device fabrication process including reactive-ion etching (RIE) and metal deposition (Supplementary \ref{sec:FTIR_fab}). We find that the spectral peaks do not shift during these procedures (except the slight shift of transition 2 for 1.67\degree\ device, see Supplementary \ref{sec:FTIR_fab}), allowing us to connect the results of quantum transport with FTIR measurements.

\section{Modelling the band structure of TBG}

In order to understand the observed optical spectra, we calculate the band structure of TBG using the continuum model\cite{bistritzer2011moire, tarnopolsky2019origin} and compare our measured spectra with optical conductivity calculations. When two layers of monolayer graphene are stacked together with a small twist angle, a long-period moiré superlattice forms (Fig. \ref{fig:Fig_1}a inset), which consists of AA and AB stacking regions. The effective Hamiltonian is constructed by considering the intralayer Hamiltonian of each layer of graphene and the effective interlayer coupling matrix between the two layers. The model incorporates several key parameters \cite{bistritzer2011moire, tarnopolsky2019origin}: (1) the twist angle, which defines the periodicity of the moiré superlattice; (2) the intralayer tunneling potential ($t$), which defines the nearest hopping energy within each graphene layer and can be readily converted to the Fermi velocity of monolayer graphene by $v_F$=$\sqrt{3}ta_0/2$ ($a_0$=0.246 nm, lattice constant of graphene). This determines the intralayer Hamiltonian; (3) the interlayer coupling energy at AA ($u_0$) and AB ($u_1$)   sites, which describes the interlayer coupling strength between twisted graphene layers. 

%%%%%%%%%%%%%%%%%%%%%%%%%%%%%%%%%%%%%
%
%%%%%%%%%%%%%%%%%%%%%%%%%%%%%%%%%%%%%
Here, we emphasize several key points about how the above-mentioned parameters modify the band structure of TBG and, consequently, the optical conductivity and band gap. 
First, the continuum model defines a magic angle 1.1\degree\ when the Fermi velocity of the flat band reaches zero\cite{bistritzer2011moire} (Supplementary \ref{sec:continuum} Fig. \ref{fig:SI_Alan}a,b) and electron correlations are expected to be the strongest. The dimensionless parameter $\alpha$ is defined as $\alpha=u_1/v_Fk_{\theta}$\cite{bistritzer2011moire,tarnopolsky2019origin}, where $k_{\theta}$ is the wave vector of the moiré superlattice and $\alpha \approx$ 0.586 for the first magic angle. Hence, the magic angle is fixed by the ratio $u_1/v_F$. 
Second, the ratio $u_0/u_1$ describes the relative size of the AA and AB regions within the moiré unit cell, which can be modified due to the lattice relaxation that favors the shrinking of the AA regions and the expansion of AB regions\cite{nam2017lattice,yoo2019atomic,bennett2023twisted}. Consequently, the optical conductivity will change both in peak energies and intensities by varying $u_0/u_1$ (Supplementary \ref{sec:continuum} Fig. \ref{fig:SI_Alan_1p46}c), which can then be compared to FTIR measurements. Moreover, the bandgaps between the flat and remote band increase with decreasing $u_0/u_1$, which can be directly compared to those extracted from electrical transport measurements (Supplementary \ref{sec:continuum} Fig. \ref{fig:SI_Alan_1p46}d). 
Third, we tentatively choose $v_F$ to be in the range of 0.89-1.2$\times$10$^6$ m/s, inclusive of the values from literature about monolayer graphene\cite{bistritzer2011moire,kim2012direct,park2021flavour}. Tuning $v_F$ shifts the energy of interband transitions monotonically without changing its intensity (Supplementary \ref{sec:continuum} Fig. \ref{fig:SI_Alan_1p46}b), and therefore $v_F$ can be extracted by comparing optical conductivity calculations with FTIR measurements (Fig. \ref{fig:Fig_3}a).  
The above considerations place strict constraints on the choice of parameters, allowing us to identify the best agreement with the experimental data while maintaining physical meaning.

We note that our model does not consider the effects of electron-electron interactions and the influence of the hBN substrate. The former is further justified by low-temperature FTIR measurements, which show no difference in the extinction spectra between 4.5\,K and 300\,K (Supplementary \ref{sec:cryoFTIR}). Regarding the substrate, we check the alignment angles between TBG and hBN crystal axes from the optical images and find alignment angles larger than 3\degree, which we expect to have little influence on the infrared spectra measured in this study (Supplementary \ref{sec:Alignment_hBN}) \cite{cea2020band}.

%%%%%%%%%%%%%%%%%%%%%%%%%%%%%%%%%%%%%
%

We now compare the continuum model calculations with experimental data for the 1.46\degree\  sample. We find the best match for $v_F$=1.05$\times$10$^6$ m/s, $u_1$=130 meV, and $u_0$=100 meV (see Supplementary \ref{sec:continuum} for details on modeling procedures). We fix $u_1/v_F$ such that the magic angle is defined for 1.1\degree, consistent with previous experiments\cite{saito2020independent,cao2021nematicity}. Fig. \ref{fig:Fig_2}d shows the calculated band structure from the continuum model with the above-mentioned parameters. Fig. \ref{fig:Fig_2}c plots the optical conductivity, which displays four prominent features like the measured spectrum, with the peak positions matching reasonably well for peaks 2-4. From this analysis, we find that transitions between the flat and first remote band  (transition 1) have the highest oscillator strength, although our spectrum is smeared by the hBN phonon band. To identify the physical origin of the peaks, we calculate the optical conductivity that only acounts for the interband transitions labeled as 1-4 in Fig. \ref{fig:Fig_2}d (Fig. \ref{fig:Fig_2}c dashed lines),  whose energies match well with the four experimental peaks in Fig. \ref{fig:Fig_2}a. As a final check of the chosen parameters and calculated band structures, we compare the bandgaps separating the flat and remote bands with experimental values extracted from temperature-dependent transport measurements. The calculated bandgap of 39 meV is in reasonable agreement with the experimental values of 50 (46)\,meV for electron (hole) doping. 

%%%%%%%%%%%%%%%%%%%%%%%%%%%%%%%%%%%%%
%

We proceed by comparing the continuum model calculations with experimental data for other devices following a similar procedure (Supplementary \ref{sec:continuum}). The experimental extinction spectra of all the devices are shown in Fig. \ref{fig:Fig_3}a, accompanied by the theoretical spectra (Fig. \ref{fig:Fig_3}b). Like the experimental data, our calculations capture the evolution of the spectral peaks to lower energies for decreasing twist angle. For twist angles 0.94\degree-1.16\degree, we find that transitions 1 and 2 shift outside the range of FTIR measurements but transitions 3 and 4 remain. Fig. \ref{fig:Fig_3}d plots a comparison of the peak positions obtained from experiments and calculations (extracted from the minima of the second derivative of the spectra, Supplementary \ref{sec:peak_extraction}), highlighting the good agreement between them.

Fig. \ref{fig:Fig_4}a plots the bandgaps extracted for the electron and hole sides, compared with those from calculations, where we obtain a reasonable agreement. The experimental data captures the trend observed previously in which the bandgap exhibits a non-monotonic dependence, reaching its maximal value of around 1.1\degree. This behavior is another hint at the magic angle for which the bandgap is supposed to be the largest\cite{carr2019exact}. 

\section{Discussion}

The continuum model parameters used to obtain the best agreement to the experimental data (Fig. \ref{fig:Fig_3}b,d and  Fig. \ref{fig:Fig_4}a) are plotted in Fig. \ref{fig:Fig_4}b as a function of our devices’ twist angle (see Supplementary \ref{sec:continuum} Table \ref{Table:parameters} for specific values). For comparison, the black and blue dashed lines plot the original continuum model parameters taken from refs\cite{bistritzer2011moire,tarnopolsky2019origin}: $v_F$=0.89 $\times$ 10$^6$ m/s, and  $u_0$=$u_1$=110 meV. Our experimental results suggest several differences compared to the original continuum model. First, the intralayer/interlayer coupling energies are not constant; instead, they vary with the twist angle. Our measurements in devices 1.12\degree-1.67\degree\  can be described by fixed $u_1$ and $v_F$. However, calculations using these parameters for 0.94\degree\  and 0.98\degree\  fail to capture the spectral shape of FTIR measurements(Supplementary \ref{sec:continuum} Fig. \ref{fig:SI_Alan_0p94}a,b). To accurately represent this spectral shape, it is necessary to vary $u_1$ and $v_F$ accordingly (Fig. \ref{fig:Fig_4}b). Second, $v_F$ is clearly underestimated by the original model. For twist angle 1.12\degree-1.67\degree, we find that $v_F$ = $1.05\times10^6$\,m/s, consistent with values obtained in a recent spectroscopic measurement of monolayer graphene\cite{inbar2023quantum}. Third, we find $u_0 < u_1$, which implies the presence of lattice relaxation, well established in the literature \cite{nam2017lattice, yoo2019atomic, carr2019exact}.

Importantly, our analysis suggests that the magic angle is not a fixed value, but rather defined over an angular range\degree\cite{carr2019exact,bennett2023twisted}. For 1.12\degree-1.67\degree, we achieve good agreement between experiments and calculations by fixing the magic angle to 1.1\degree\cite{bistritzer2011moire,tarnopolsky2019origin}, consistent with the observation of the highest superconducting critical temperature and the largest correlated insulator bandgap $\nu$ = -2 at that angle \cite{saito2020independent, cao2021nematicity}. However, the calculated spectra of 0.94\degree\  and 0.98\degree\   do not match the experimental data under the same choice of parameters, specifically, those satisfying the condition where the magic angle is equal to 1.1\degree\ (Supplementary \ref{sec:continuum} Fig. \ref{fig:SI_Alan_0p94}a,b). Theoretical work\cite{carr2019exact,bennett2023twisted} suggests that the magic angle may shift to a lower angle considering the effect of the lattice relaxation, instead of being a single fixed value. Inspired by this prediction, we achieve good agreement for 0.94\degree\  and 0.98\degree\  when the magic angle of the continuum model is shifted from 1.1\degree\ to $\sim$0.9\degree\ (Supplementary \ref{sec:continuum} Fig. \ref{fig:SI_Alan_0p94} and \ref{fig:SI_Alan_0p94_vf} ). This is also compatible with our transport measurements that reveal correlated features, intrinsic to magic-angle TBG in 0.94\degree\  and 0.98\degree\ devices (Fig. \ref{fig:Fig_3}c, Supplementary \ref{sec:transport}). Several experimental works have also shown correlated states and superconductivity in samples with twist angles between 0.9\degree-1.0\degree\cite{choi2019electronic,cao2021nematicity}, further supporting this claim.

\section{Conclusion and Outlook}
In conclusion, we present a systematic optical spectroscopy study on TBG. We measure the optical transitions unique to TBG and connect them with the bandgaps extracted from quantum transport in a single heterostructure. Although we employ the continuum model in the simplest form with only essential parameters, this study provides a starting point for a quantitative understanding of the single-particle band structure, which can be used for the development of further comprehensive theoretical models. Apart from this study, terahertz optical spectroscopy holds huge potential for probing the strongly correlated electrons inside the flat band\cite{yang2022spectroscopy}.

Finally, we recall that the infrared spectra are extremely sensitive to the twist angle, which can be measured at room temperature by FTIR directly after heterostructure assembly. The technique therefore may be used as a pre-characterization tool before nanofabrication and cryogenic measurements.

\begin{figure*}[]
    \centering
    \includegraphics[width=0.8\textwidth]{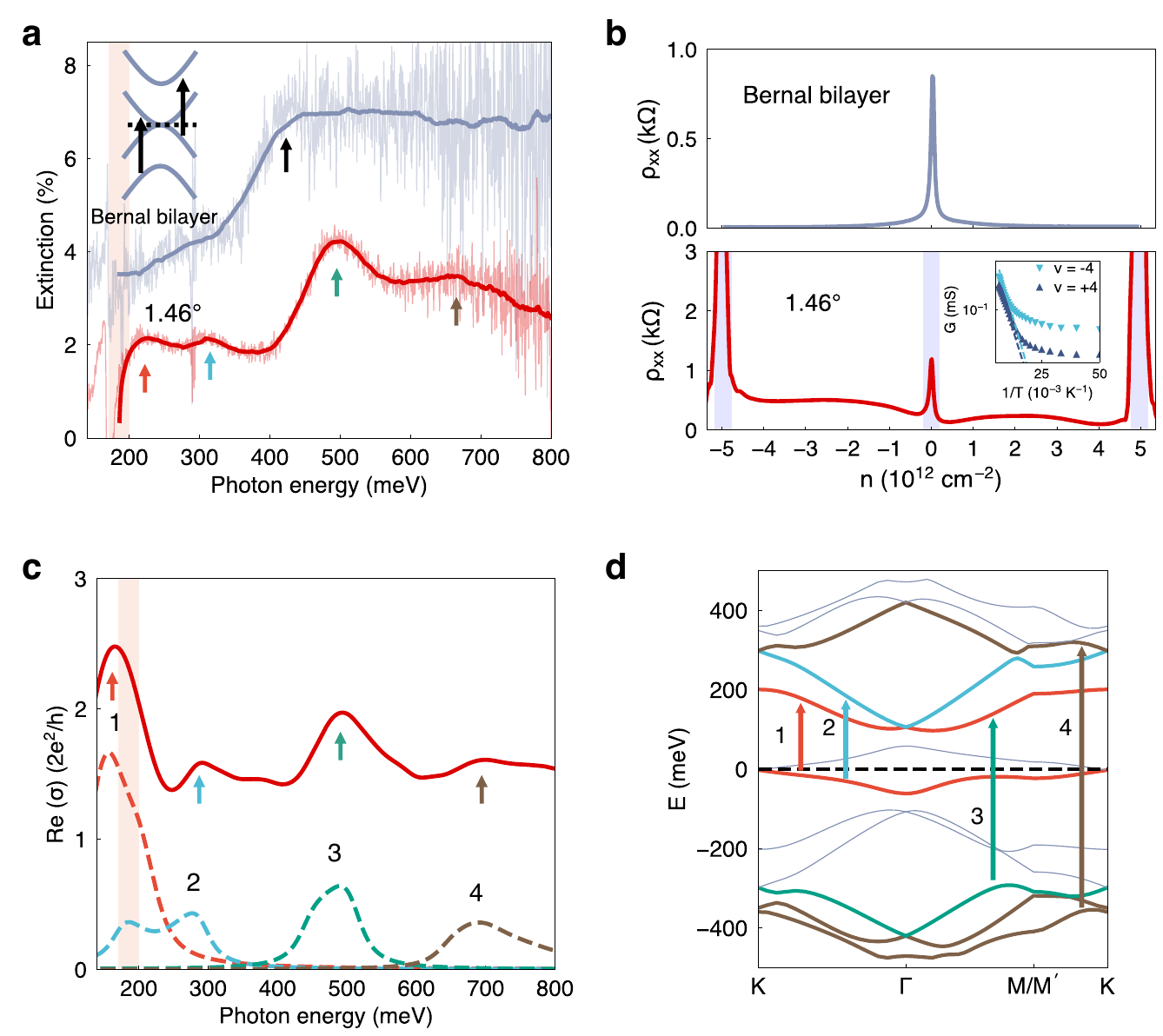}
    \caption{\textbf{TBG 1.46\degree. }\textbf{a,} Extinction spectra (raw and smoothed) of  Bernal bilayer graphene and 1.46\degree\  TBG (offset by 2\%\ for clarity). The orange-shaded region indicates the Reststrahlen phonon band of hBN. Inset: Band structure of Bernal bilayer graphene. The dashed line is the Fermi level.  The black arrows indicate interband transitions responsible for optical absorptions close to 400 meV. The colored arrows label absorptions peaks of 1.46\degree\ corresponding to the interband transitions as indicated in \textbf{d}. \textbf{b,} Resistivity versus carrier density of Bernal bilayer graphene and 1.46\degree\  TBG. Inset: Conductance at $\pm$ 4 fillings (triangles) fitted by exp(-$\Delta$/2$k_B$T) (dashed lines), where $\Delta$ is the bandgap and $k_B$ is the Boltzmann constant.  \textbf{c.} The real part of the optical conductivity Re($\sigma$) computed from the band structure in \textbf{d} using the Kubo formula. The dashed lines are Re ($\sigma$) that only count for the transitions labeled by the arrows 1-4 in \textbf{d}. \textbf{a} and \textbf{c} share the same x-axis. \textbf{d,} Band structure of 1.46\degree\ derived from the continuum model as discussed in the main text. The arrows label the optical transitions we observe in the extinction spectrum as displayed in \textbf{a}. }
    \label{fig:Fig_2}     
\end{figure*}
 
\begin{figure*}[]
 
    \includegraphics[width=1\textwidth]{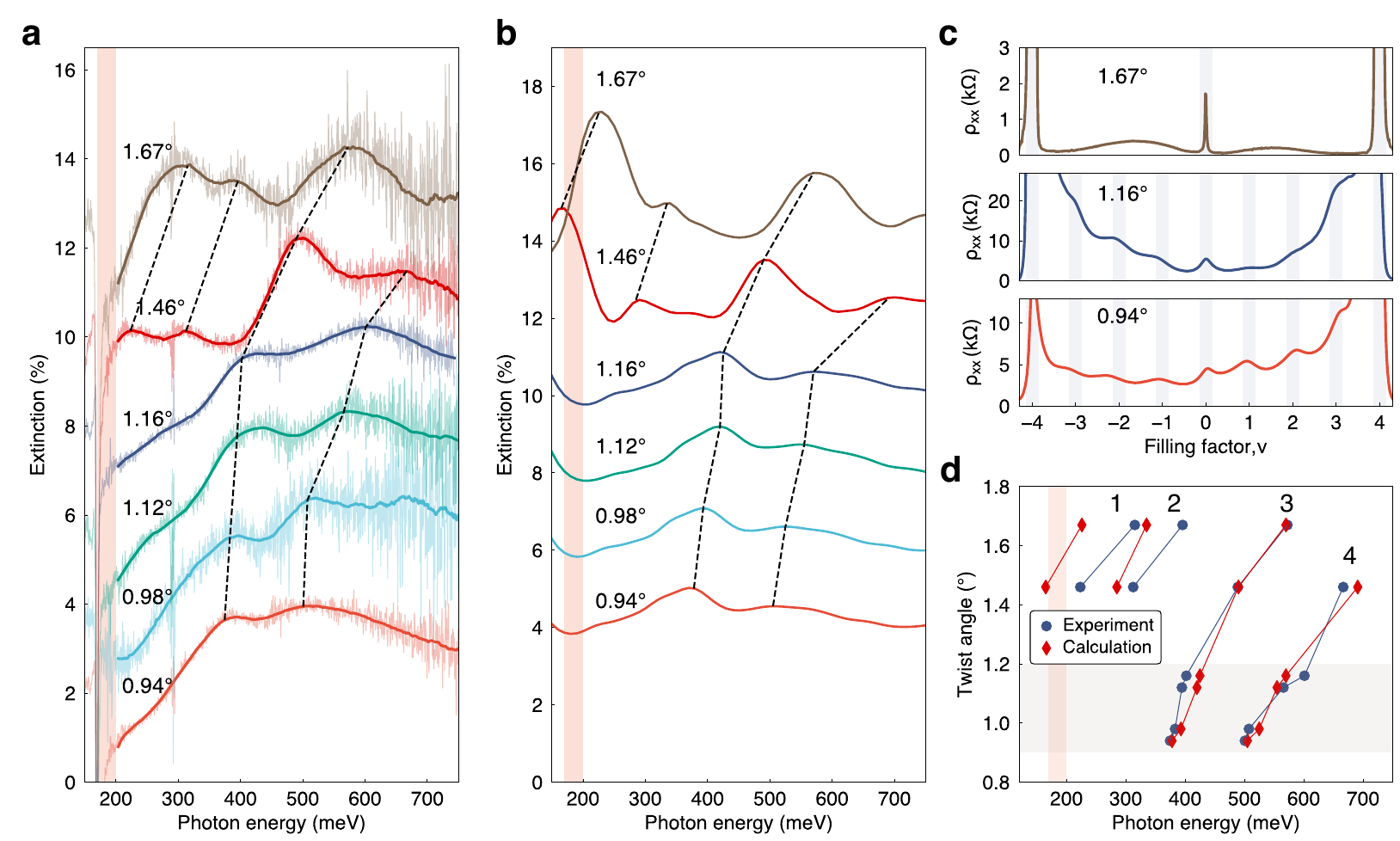}
    \caption{\textbf{Twist angle dependent extinction spectra. }\textbf{a,} Extinction spectra (raw and smoothed) of TBG with twist angles between 0.94\degree-1.67\degree. The black dashed lines indicate the evolvement of the absorption peaks with the twist angle. \textbf{b,} Extinction spectra of TBG computed by transfer matrix method with the input of optical conductivity from the continuum model, which can be directly compared with the spectra in \textbf{a}. The spectra in \textbf{a} and \textbf{b} are offset for clarity. The orange-shaded region indicates the Reststrahlen phonon band of hBN. \textbf{c,} Resistivity versus filling factor for  0.94\degree, 1.16\degree, and 1.67\degree. \textbf{d,} Optical absorption peaks versus twist angle extracted from the experimental (\textbf{a}) and calculated (\textbf{b}) optical spectra by finding the minima of the second derivative (Supplementary \ref{sec:peak_extraction}). The shaded region (0.9\degree-1.2\degree) indicates the angular range where correlated insulating states and superconductivity are observed in transport measurements. The solid lines are guidance for the eyes. Transitions 1-4 correspond to the optical transitions labeled in Fig \ref{fig:Fig_2} \textbf{d}. }
    \label{fig:Fig_3}
\end{figure*}
 
\begin{figure*}[]
 
    \includegraphics[width=0.9\textwidth]{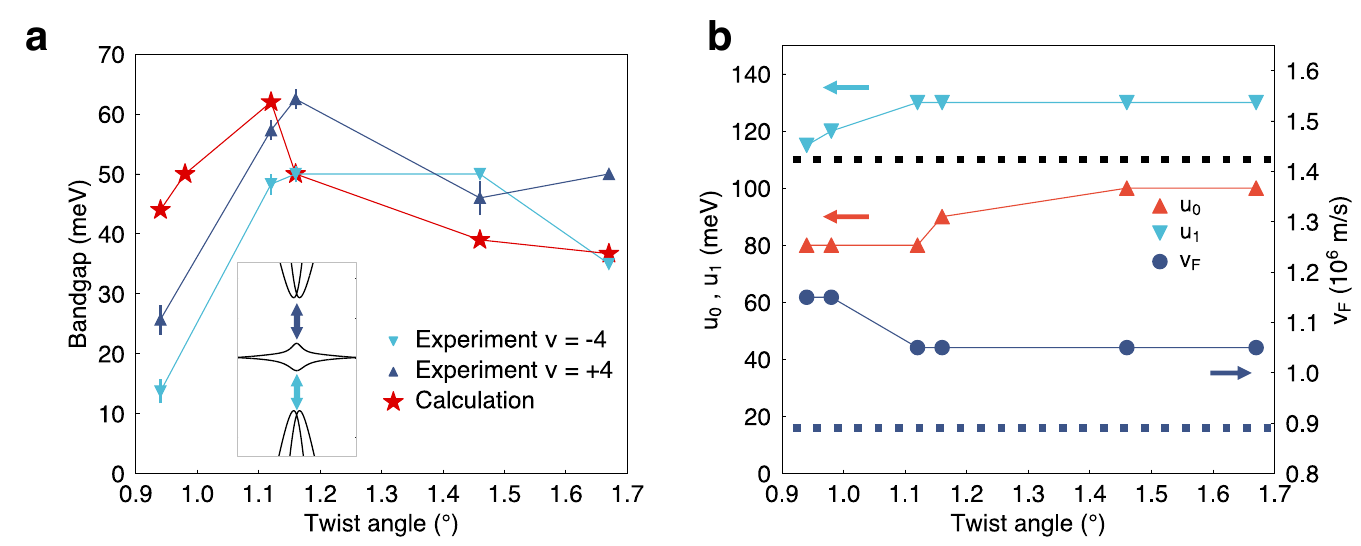}
    \caption{ \textbf{Bandgap at $\nu$=$\pm$4 and the continuum model parameters.} \textbf{a,} Bandgap at $\nu = \pm 4$ for TBG devices with twist angles between 0.94\degree\ and 1.67\degree. Error bars indicate the standard deviation of the bandgaps extracted from different pairs of contacts. The red stars are the bandgap at $\nu = \pm 4$ calculated with the continuum model using the parameters in \textbf{b}. \textbf{b,} Interlayer coupling energies (left y-axis) at AA site ($u_0$, red) and AB site ($u_1$, cyan) and Fermi velocity of monolayer graphene ($v_F$, dark blue) (right y-axis) as a function of the twist angle obtained by comparing the continuum model calculations to FTIR and transport measurements (see Supplementary \ref{sec:continuum} for details). As a comparison to the previous theoretical work, the black dashed line is $u_0$=$u_1$=110 meV\cite{bistritzer2011moire,tarnopolsky2019origin}, and the dark blue dashed line is $v_F$=0.89$\times$$10^6$ m/s\cite{bistritzer2011moire,tarnopolsky2019origin}.  }
    \label{fig:Fig_4}
\end{figure*}

\clearpage
%
%%%%%%%%%%%%%%%%%%%%%%%%%%%%%%%%%%%%%%%
\section{Declarations}
\subsection{Acknowledgements}
% reordered in line with the author list

We thank Valerio Pruneri for the use of FTIR. We acknowledge discussions with Cheng Shen and Hanan Herzig Sheinfux, as well as experimental assistance from Karuppasamy Soundarapandian and Matteo Ceccanti. The research leading to these results has received funding from the European Union’s Horizon 2020  under grant agreement no. 881603 (Graphene flagship Core3) and 820378 (Quantum flagship).
R.K.K. acknowledges the EU Horizon 2020 program under the Marie Skłodowska-Curie grants 754510 and 893030 and the FLAG-ERA grant (PhotoTBG), by ICFO, RWTH Aachen and ETHZ/Department of Physics.
P.S. acknowledges support from the European Union’s Horizon 2020 research and innovation program under the Marie Skłodowska-Curie Grant No. 754510.
P.A.P and F.G. acknowledge funding from the European Commission, within the Graphene Flagship, Core 3, grant number 881603 and from grants NMAT2D (Comunidad de Madrid, Spain), SprQuMat and (MAD2D-CM)-MRR MATERIALES AVANZADOS-IMDEA-NC, with support from the \textquotedblleft Severo Ochoa\textquotedblright~Programme for Centres of Excellence in R\&D (CEX2020-001039-S/AEI/10.13039/501100011033).
Z.Z. acknowledges support funding from the European Union's Horizon 2020 research and innovation program under the Marie Skłodowska-Curie grant agreement No 101034431.
H.A. also acknowledges funding from the European Union’s Horizon 2020 research and innovation program under the Marie Skłodowska-Curie grant agreement No. 665884.
A.B. and A.B.K. acknowledge support from the Swiss National Science Foundation.
J.B. acknowledges support from the European Union’s Horizon Europe program under grant agreement 101105218.
K.W. and T.T. acknowledge support from the JSPS KAKENHI (Grant Numbers 21H05233 and 23H02052) and World Premier International Research Center Initiative (WPI), MEXT, Japan.
F.H.L.K. acknowledges support from the ERC TOPONANOP (726001), the government of Spain (PID2019-106875GB-I00; Severo Ochoa CEX2019-000910-S [MCIN/ AEI/10.13039/501100011033], PCI2021-122020-2A funded by MCIN/AEI/10.13039/501100011033), the "European Union NextGenerationEU/PRTR (PRTR-C17.I1), Fundació Cellex, Fundació Mir-Puig, and Generalitat de Catalunya (CERCA, AGAUR, 2021 SGR 01443).
This material is based upon work supported by the Air Force Office of Scientific Research under award number FA8655-23-1-7047.
Any opinions, findings, and conclusions or recommendations expressed in this material are those of the author(s) and do not necessarily reflect the views of the United States Air Force.

\subsection{Author contributions}

F.H.L.K. conceived the project. F.H.L.K. and R.K.K. supervised the project. G.L. fabricated the devices under the guidance of P.S. and H.A. G.L. designed and performed the FTIR measurements. G.L. performed the transport measurements on a system built by R.K.K. with the help of R.K.K., P.S., H.A. and J.B. I.T. developed the Python package for the continuum model. G.L. performed the calculations with I.T.'s Python package under the supervision of I.T., R.K.K., and F.H.L.K. P.A.P. and Z.Z. performed the calculations under the supervision of F.G. A.B. carried out the cryogenic FTIR measurements under the supervision of A.B.K. The hBN crystals are provided by K.W. and T.T.  G.L., R.K.K., J.B. and F.H.L.K. wrote the manuscript with input from all authors.

\subsection{Competing Financial Interests}

The authors declare no competing financial interests.

\subsection {Availability Statement}

The data that supports the plots within this paper is available at this link: \url{https://doi.org/10.5281/zenodo.10912482}.

%%%%%%%%%%%%%%%%%%%%%%%%%%%%%%%%%%%%%%%
%

\section{Methods}
\subsection{Device fabrication}
The samples are fabricated by the ``cut-and-stack'' method\cite{stepanov2020untying,saito2020independent}. A polydimethylsiloxane (PDMS) stamp covered with a propylene carbonate (PC) film is used to pick up the top hBN flake. Graphene flake is mechanically exfoliated on  Si{\footnotesize ++}/SiO{\footnotesize 2} (285 nm) and cut into two pieces using an AFM tip. The top hBN is then used to pick up the first graphene flake followed by the second graphene with a target rotation angle ($\sim$1.15$\degree$ for magic-angle devices). The heterostructure then picks up bottom hBN and few-layer graphite as a gating electrode. During the transfer process, the substrate is kept at 100-120 \celsius. The final stack is then dropped on a target substrate at 180 \celsius. The PC film is them removed by chloroform. The hBN flakes and graphite gate are carefully chosen to have uniform thicknesses without layer steps. The thickness of hBN is usually 10-20 nm. We use a graphite gate instead of a metal gate for two reasons: 1. Although few-layer graphite has infrared absorptions, it is still able to transmit most of the infrared light and therefore makes it possible to perform transmission measurements; 2. Graphite gate is proven to yield a high device quality thanks to the clean interface between graphite and hBN\cite{zibrov2017tunable, stepanov2020untying,lu2019superconductors}. We use lightly doped silicon (0.5 mm thickness, P-type, B-doped, 2 sides polished, 1-20 ohm.cm) with double-side coated SiO{\footnotesize 2} (300 nm each side) as the device substrate, which is semi-transparent in the infrared frequency range. The stack is etched into a hall bar geometry by SF{\footnotesize 6} and O{\footnotesize 2} plasma etching and one-dimension contacts (Cr/Au, 5/60 nm) are deposited by electron beam evaporation (Cr) and thermal evaporation (Au).

\subsection{FTIR measurements}
The infrared spectra are collected from Bruker Tensor II with a Hyperion microscope (Objective 35X, NA=0.5) equipped with a mercury cadmium telluride (MCT) detector cooled by liquid nitrogen. The measurements are done at room temperature in the ambient atmosphere without nitrogen purging. Carbon dioxides in the air have a strong absorption at 291.3 meV (2350 cm$^{-1}$) and will cause a spike/dip in the final extinction spectrum (Fig. \ref{fig:Fig_2}a, Fig. \ref{fig:Fig_3}a), which also confirms the wavenumber accuracy of the instrument. The spectra are acquired in a rapid scan mode averaged over 600 scans with a resolution of 4 cm$^{-1}$. The aperture for the globar thermal source is 6 mm. The noise level of the final spectrum varies from 0.1\% to 2\% depending on the physical dimension of the device. The spectrum at the reference region is collected immediately after the signal region while the FTIR is kept at exactly the same condition and the scanning parameters are also the same. We chose transmission measurement instead of reflection because transmission mode gives a better SNR compared to reflection in our measurements. We did not apply electrostatic gating to the device during the FTIR measurements and thus the doping level was close to zero, which was later confirmed by transport measurements that the gate voltage of the charge neutrality point was always close to zero.

\subsection{Transport measurements}

Electrical measurements were carried out using the usual 4-probe transport techniques in a dry cryostat. Briefly, we source small excitation currents (10-100 nA) at low frequency (17.111 Hz) through the device's leads. Lock-in techniques are used to measure the voltage across different pairs of contacts. A Keithley 2614B source meter is used to control the gate voltage of the device. The carrier density is calibrated using Hall measurements under out-of-plane magnetic fields of $\pm$ 0.95 T.

%%%%%%%%%%%%%%%%%%%%%%%%%%%%%%%%%%%%%%%
%
\makeatletter
\renewcommand{\thefigure}{S\arabic{figure}}
\makeatother
\setcounter{figure}{0} % reset counter

\section{Supplementary Information}
\label{sec:SI}
\subsection{Devices homogeneity}
\label{sec:angle_inhomogeneity}

\begin{figure*}[h]
    \centering
    \includegraphics[width=0.9\textwidth]{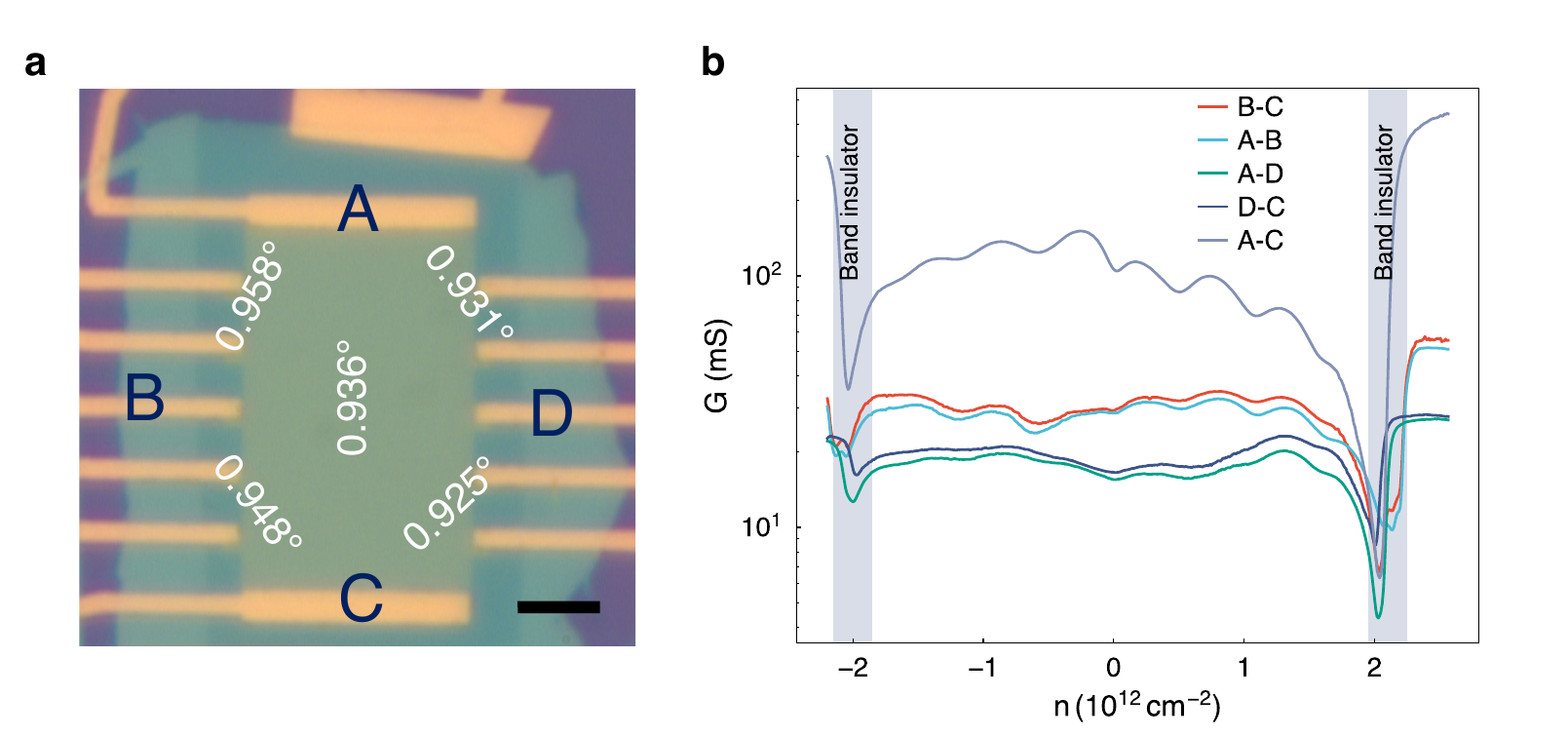}
    \caption{\textbf{Two terminal conductance measurements.}  \textbf{a,} Optical image of a 0.94 $\degree$ device after fabrication. Scale bar, 5 \textmu m. The numbers labeled on the image are the measured twist angles between different pairs of contacts (extracted from the maxima of resistance at $\pm$4 fillings). \textbf{b,} Two terminal conductance measurements from different pairs of contacts. Data were taken at 10\,K. }
    \label{fig:SI_inhomegenity}  
\end{figure*}

Fig. \ref{fig:Fig_3}a of the main text shows that optical transitions 1-4 of TBG are sensitive to the twist angle. For instance, we observe a clear spectral shift between 1.12\degree\  and 1.16\degree. Therefore, the twist-angle homogeneity of our devices is critical. We employ high temperature ($\sim$100\degree C) pickup during the heterostructure assembly to remove bubbles and yield high-quality devices. This is seen in atomic force microscope (AFM) images (Fig. \ref{fig:SI_alignment} b). The twist-angle homogeneity is checked on all our devices using two-probe conductance measurements between all pairs of contacts (Fig. \ref{fig:SI_inhomegenity}). The twist angle variations are limited to $\pm 0.02$\degree\ in all our devices.

%

%%%%%%%%%%%%%%%%%%%%%%%%%%%%%%%%%%%%%%%
\subsection{Twist angle extraction}
\label{sec:angle_extraction}

\begin{figure*}[h]
    \centering
    \includegraphics[width=0.9\textwidth]{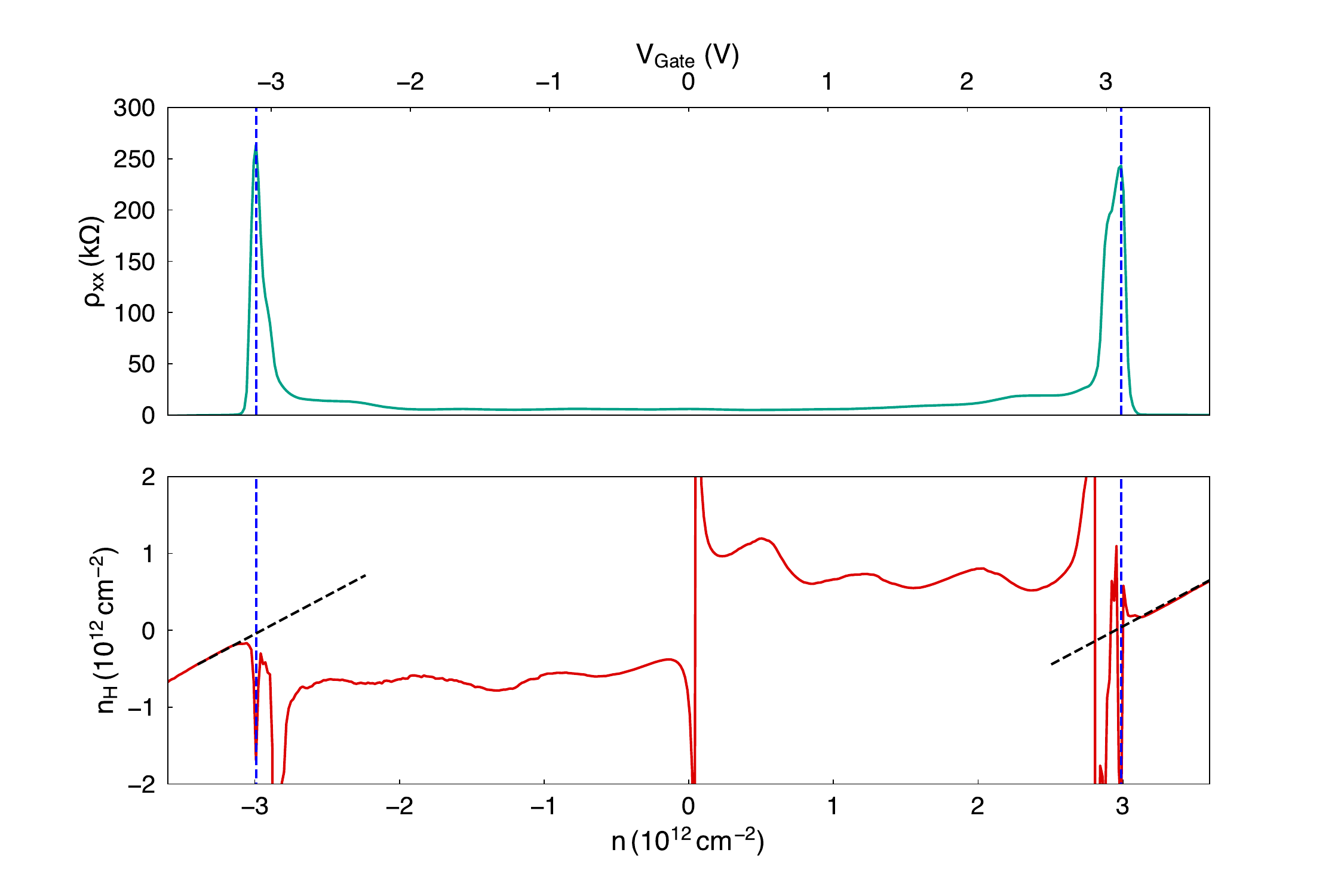}
    \caption{\textbf{Hall density measurement of a 1.12$\degree$ TBG.} Four-probe resistivity ($\rho_{xx}$) (top) and hall density measurement ($\rho_{xy}$) (bottom) under a magnetic field of 0.95 T. Black dashed lines: fitting with a planar capacitor model in the region where the classical regime is applicable. Blue dashed lines indicate the resistivity maxima at $\nu = \pm 4$, allowing calculation of the twist angle.}
    \label{fig:SI_Rxy}  
\end{figure*}

We extract the twist angles by using 4-terminal transport measurements. Fig. \ref{fig:SI_Rxy}a shows the typical resistivity for our devices. Two peaks were systematically measured, corresponding to the regime where the Fermi energy reaches the band insulator. In such condition, the band is fully filled with 4 (2 spin and 2 valleys) charge carriers per moiré unit cell. The twist angle is later found using the relation $n_{\nu=\pm 4} = 8 \theta / \sqrt{3} a^2$, where $a = 2.46$\,\AA\, is graphene’s lattice constant and $n$ the carrier density.
The latter was extracted from measurements of the Hall resistivity in regions of the phase diagram where the classical regime ($R_{xy} = B / |n| e$) is applicable. In our devices, this can be achieved for magnetic fields $B = \pm$ 0.95\,T and gate voltages $|V_g| > |V_{\nu = \pm 4}|$.  Fig. \ref{fig:SI_Rxy}b shows the Hall density as a function of gate voltages. We can also extract the relative permittivity of hBN ($\varepsilon_{hBN}$) through the equation: $C=\varepsilon_0\varepsilon_{hBN} /d$, where $d$ is the thickness of bottom hBN measured by AFM, and $\varepsilon_0$ is the vacuum permittivity. We extract $\varepsilon_{hBN}$=2.8 for all the devices (except 0.98\degree\ due to gate leakage) as shown in the main text.

\subsection{Additional transport data}
\label{sec:transport}
\noindent

\begin{figure*}[h]
    \centering
    \includegraphics[width=0.9\textwidth]{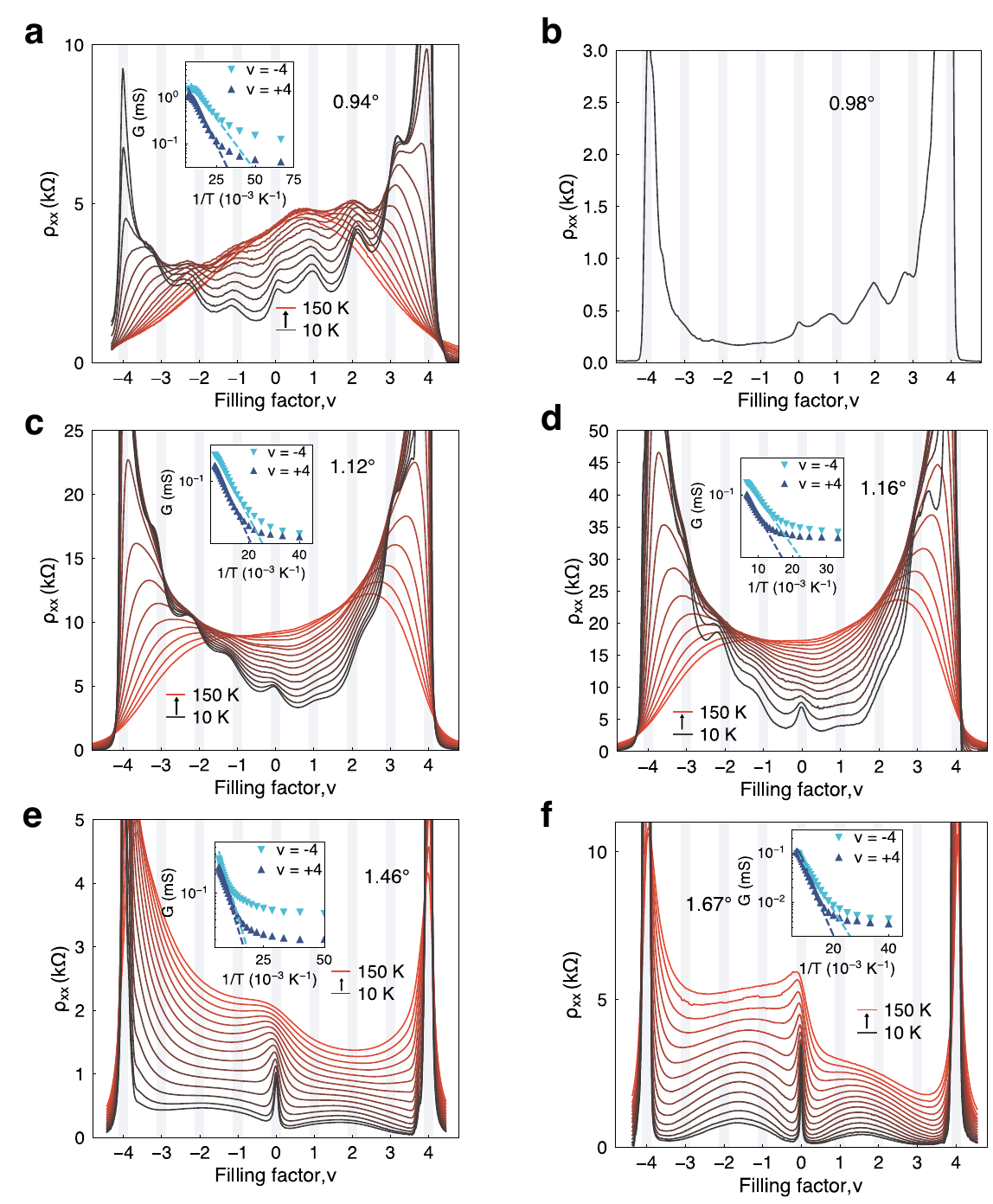}
    \caption{\textbf{Additional transport data.} Temperature-dependent measurements (10-150\,K) for all devices shown in the main text. Insets: Arrhenius fitting at $\nu=\pm 4$. For the 0.98\degree\ device, we only show data at 10\,K because of gate leakage.}
    \label{fig:SI_transport}  
\end{figure*}

In this section, we show temperature-dependent transport data for all the devices presented in the main text and related Arrhenius fitting (Fig. \ref{fig:SI_transport}), which reproduces previous transport studies. The 0.98\degree\  device started leaking before temperature-dependent and hall measurements. So we extracted the twist angle by the planar capacitor model with relative permittivity the same as extracted from all other devices ($\varepsilon_{hBN}$=2.8) and hBN thickness measured by AFM. 
%

%%%%%%%%%%%%%%%%%%%%%%%%%%%%%%%%%%%%%%%
\subsection{Alignment between graphene and hBN}
\label{sec:Alignment_hBN}

\begin{figure*}[h]
    \centering
    \includegraphics[width=0.9\textwidth]{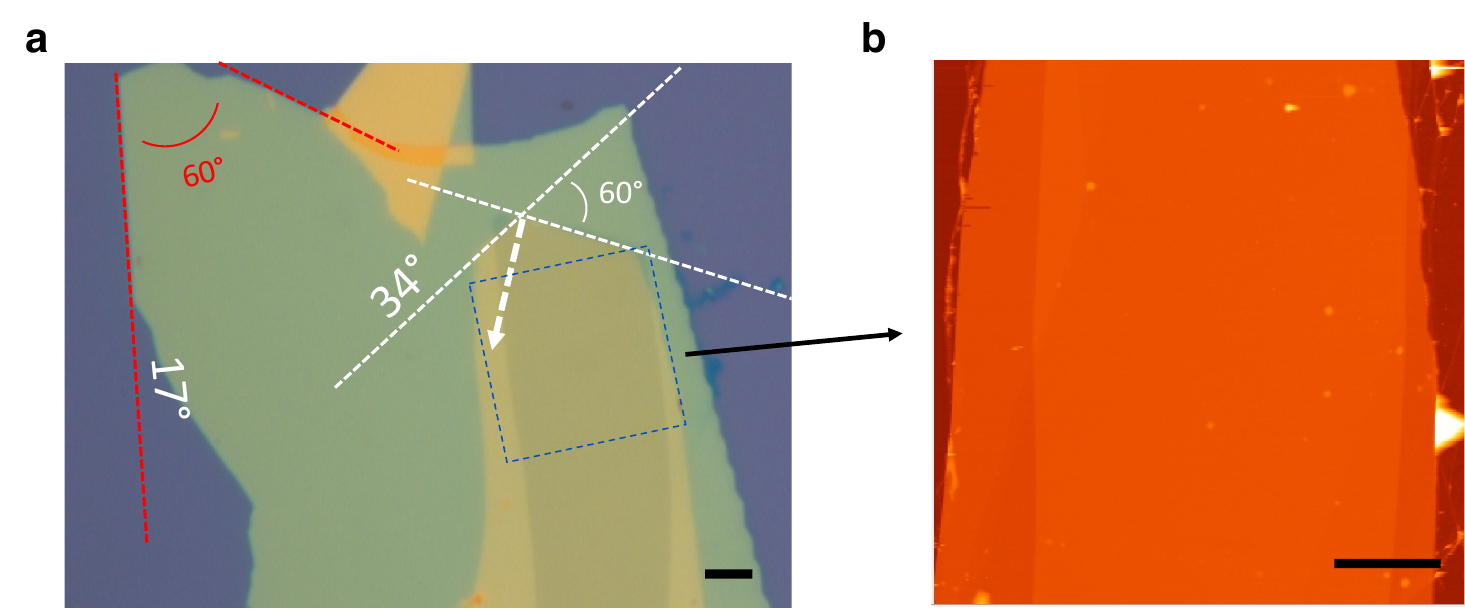}
    \caption{\textbf{Check alignment between graphene and hBN}  \textbf{a,} Optical image of the 0.94$\degree$ device. White dashed lines indicate the edge of the top hBN. Red dashed lines: edge of the bottom hBN. White arrows: edge of graphene layers. We estimate the twist angle between the top hBN and graphene to be about 4$\degree$ (or 34$\degree$, $\pm$1.5$\degree$) and between the bottom hBN and graphene to be about 17$\degree$ ($\pm$1.5$\degree$). The twist angle between the top and bottom hBN is about 9$\degree$. Scale bar, 5 \textmu m. \textbf{b,} AFM image of the twisted region as indicated by the blue dashed box in \textbf{a}. Scale bar, 5 $\mu$m.}
    \label{fig:SI_alignment}  
\end{figure*}

In the main text, the continuum model in a single-particle picture is employed considering only a set of basic parameters (twist angle $\theta$, Fermi velocity $v_F$, and interlayer coupling energies $u_0$ and $u_1$). However, due to the slight lattice mismatch between hBN and graphene, the alignment between them ($<$1\degree) can introduce a second-order moiré superlattice in addition to that of TBG, modifying the band structure and electrical properties\cite{cea2020band,serlin2020intrinsic}. We check the alignment angle between graphene and hBN by using the optical image. Fig. \ref{fig:SI_alignment} a shows an optical image of the 0.94\degree\ device with crystal edges of hBN and graphene labeled by the dashed lines. Table \ref{Table:alignment} shows that the twist angles between TBG and encapsulating hBN layers for all the devices in this study are over 3\degree. In addition, we do not observe any high resistance peaks at the charge neutrality point in all our devices (Supplementary \ref{sec:transport}), which is typically seen in TBG devices aligned to hBN\cite{serlin2020intrinsic}. Therefore, we conclude that the additional moiré superlattice may not be a dominating factor in affecting our analysis using the continuum model.

\subsection{Extinction spectra during fabrication}
\label{sec:FTIR_fab}
\noindent

In the main text, we show the extinction spectra of the stacks and transport measurements of the same stacks after a series of fabrication processes (electron beam lithography, reactive ion etching (RIE), and metal deposition). One may question whether the twist angle of the stack has changed during fabrication. Here, we show that we can monitor the optical absorptions of TBG devices during fabrication. 

\begin{figure*}[h]
    \centering
    \includegraphics[width=0.7\textwidth]{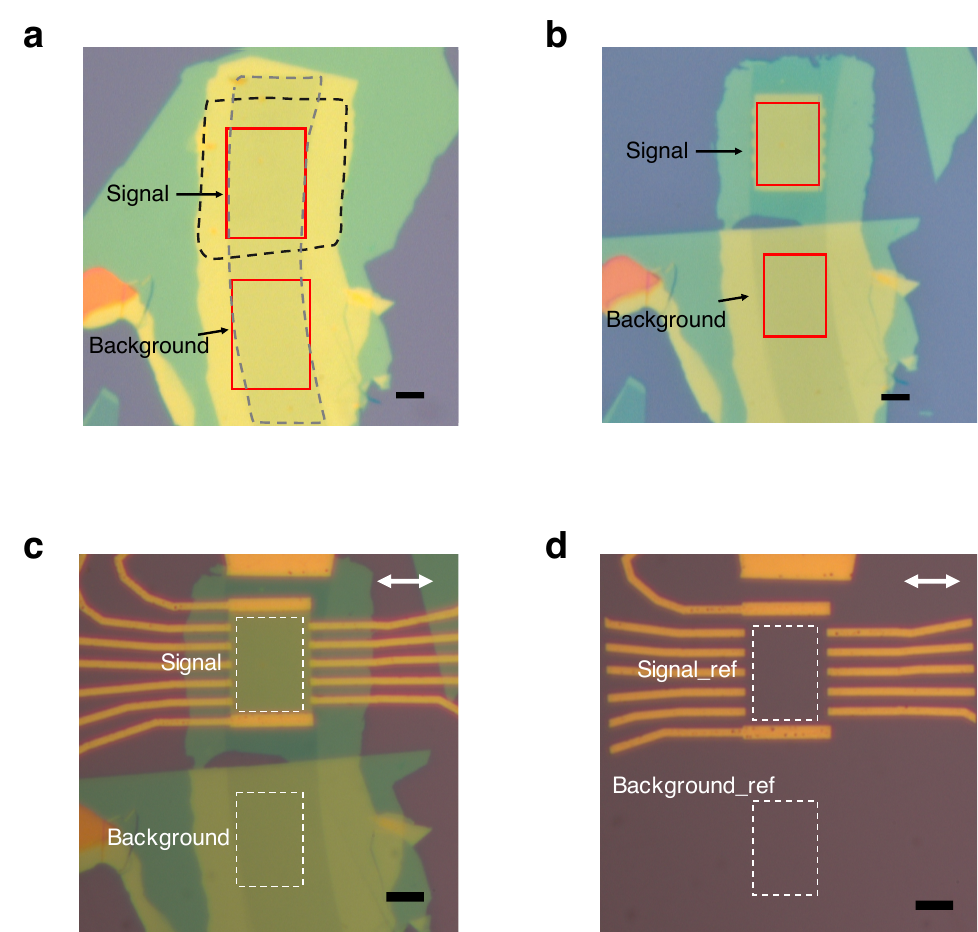}
    \caption{\textbf{Optical images for 1.12\degree\ device during fabrication.} \textbf{a.} After transfer. Black dashed line indicates the TBG region. Grey dashed line indicates the graphite gate at the bottom of the stack. Red rectangles indicate the area where the signal and background spectra are collected. \textbf{b.} After RIE. \textbf{c.} After metal deposition. White arrow indicates the polarization of the incident light. White rectangles indicate where we collected the spectra. \textbf{d.} Bare metal contacts with the same shape as \textbf{c} fabricated in order to remove the optical response of the contacts. Scale bars, 5 \textmu m in all panels.}
    \label{fig:SI_RIE_contacts_image}  
\end{figure*}

The FTIR spectra as displayed in the main text are measured from the stacks right after heterostructure assembly without any fabrication process yet (Fig. \ref{fig:SI_RIE_contacts_image}a). After RIE, the stack will be etched into hall bar geometry (Fig. \ref{fig:SI_RIE_contacts_image}b), whose extinction spectrum can be measured in a similar way as the original stack except that the device area is slightly smaller. Following, metal contacts (Cr/Au) are deposited for transport measurements (Fig. \ref{fig:SI_RIE_contacts_image}c). However, metal contacts introduce significant absorptions and reflections of the infrared light due to the Drude and plasmonic responses, which makes background normalization challenging. We design all the metal contacts to be strictly parallel to each other (Fig. \ref{fig:SI_RIE_contacts_image}c) and confine the polarization of the incident light to be parallel to the contacts by using a ZeSe holographic wire grid polarizer, minimizing the plasmonic response of the metal contacts. We also deposited the same contacts on the same substrate near the device (Fig. \ref{fig:SI_RIE_contacts_image}d), which allows us to measure the extinction spectrum of the contacts by (1 - Signal\_ref/Background\_ref)$\times$100\%. Finally, we can get the extinction spectrum of TBG by 
 [(1 - Signal/Background) - (1 - Signal\_ref/Background\_ref)]$\times$100\%. Therefore, the Drude response of the metal contacts is also properly removed. The polarizer and metal contact will decrease the transmitted signal significantly. Therefore, the spectrum after depositing contact is noisier compared to that of the original stack.

\begin{figure*}[h]
    \centering
    \includegraphics[width=0.7\textwidth]{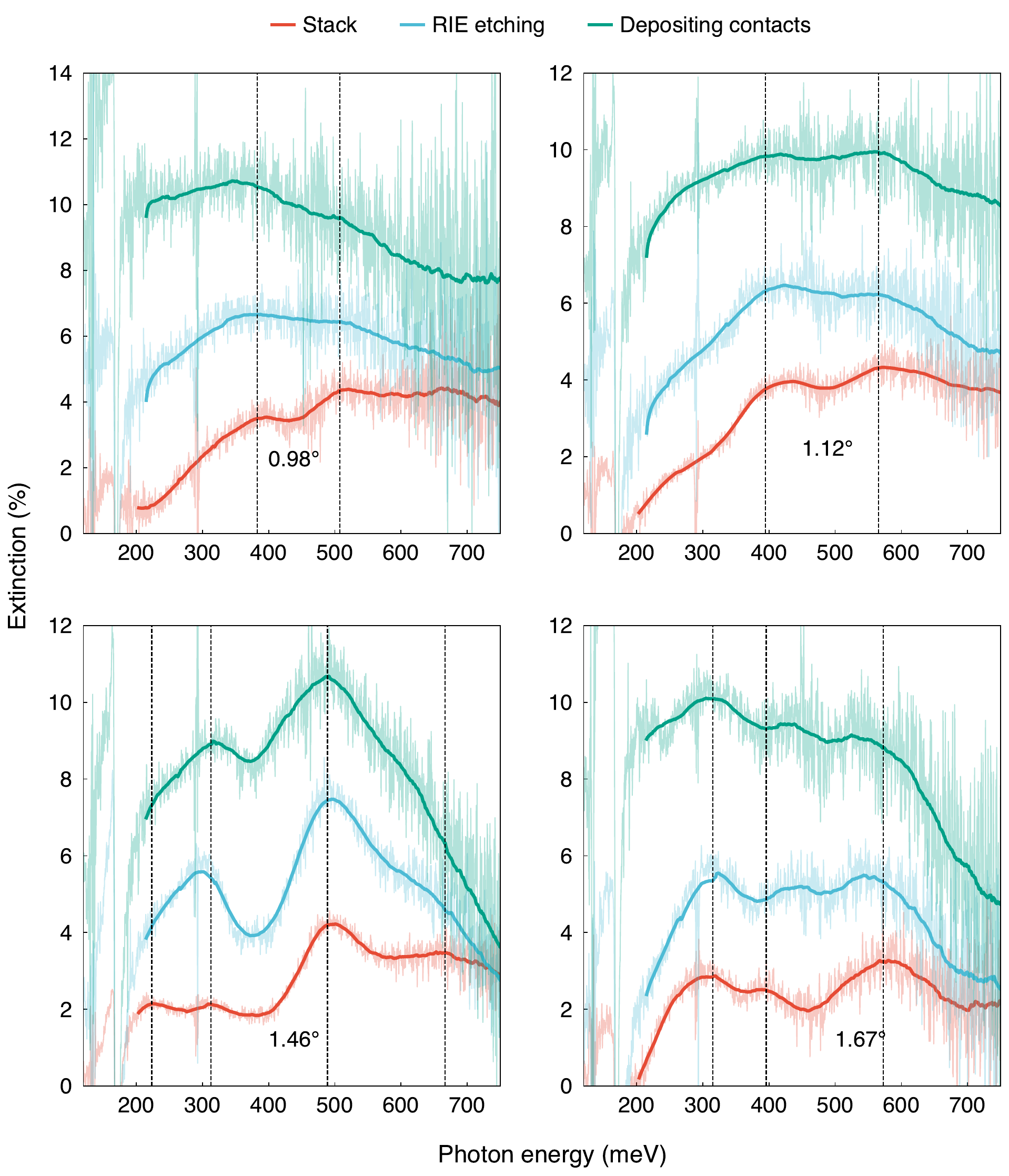}
    \caption{\textbf{Extinction spectra during fabrication} Extinction spectra of 0.98$\degree$, 1.12$\degree$, 1.46$\degree$, and 1.67$\degree$ are measured from the device after stack assembly, after RIE, and after metal deposition. The black lines indicate the peak positions extracted from the spectra of the original stack before fabrication (red line). The spectra are offset for clarity. }
    \label{fig:SI_RIE_contacts}  
\end{figure*}

Fig. \ref{fig:SI_RIE_contacts}  shows the extinction spectra of TBG with twist angles of 0.98\degree, 1.12\degree, 1.46\degree, and 1.67\degree\  at different stages: (1). the stacks after the transfer process, (2). after RIE etching but without depositing metal contacts, (3). after depositing metal contacts. The extinction spectra of 0.98$\degree$ and 1.12$\degree$ maintain almost the same shape and peak positions during fabrication. For the 1.46$\degree$ device, the peaks at 313 meV and 489 meV persist through the fabrication process, while the peaks at 223 meV and 666 meV become less clear likely because of the decreased signal-noise-ratio as discussed above.  For the 1.67$\degree$ device, the peaks at 315 meV and 572 meV maintain the same positions. However, the peak at 396 meV shifts to 430 meV after RIE etching, which we speculate may be caused by strain at the edges of the device mesa induced by etching. Nevertheless, considering all other spectral peaks remain unshifted, our results suggest that the twist angle of our devices does not change during device fabrication.

\subsection{Continuum model calculations and comparisons with experimental data}
\label{sec:continuum}
\noindent

In this section, we focus on discussing the procedure for modeling the band structure of TBG. We compare the continuum model\cite{dos2007graphene,bistritzer2011moire} calculations to the experimental data by varying $v_F$, $u_0$, and $u_1$, while sticking to the physical constraints as discussed in the main text. 

\begin{figure*}[h]
    \centering
    \includegraphics[width=0.9\textwidth]{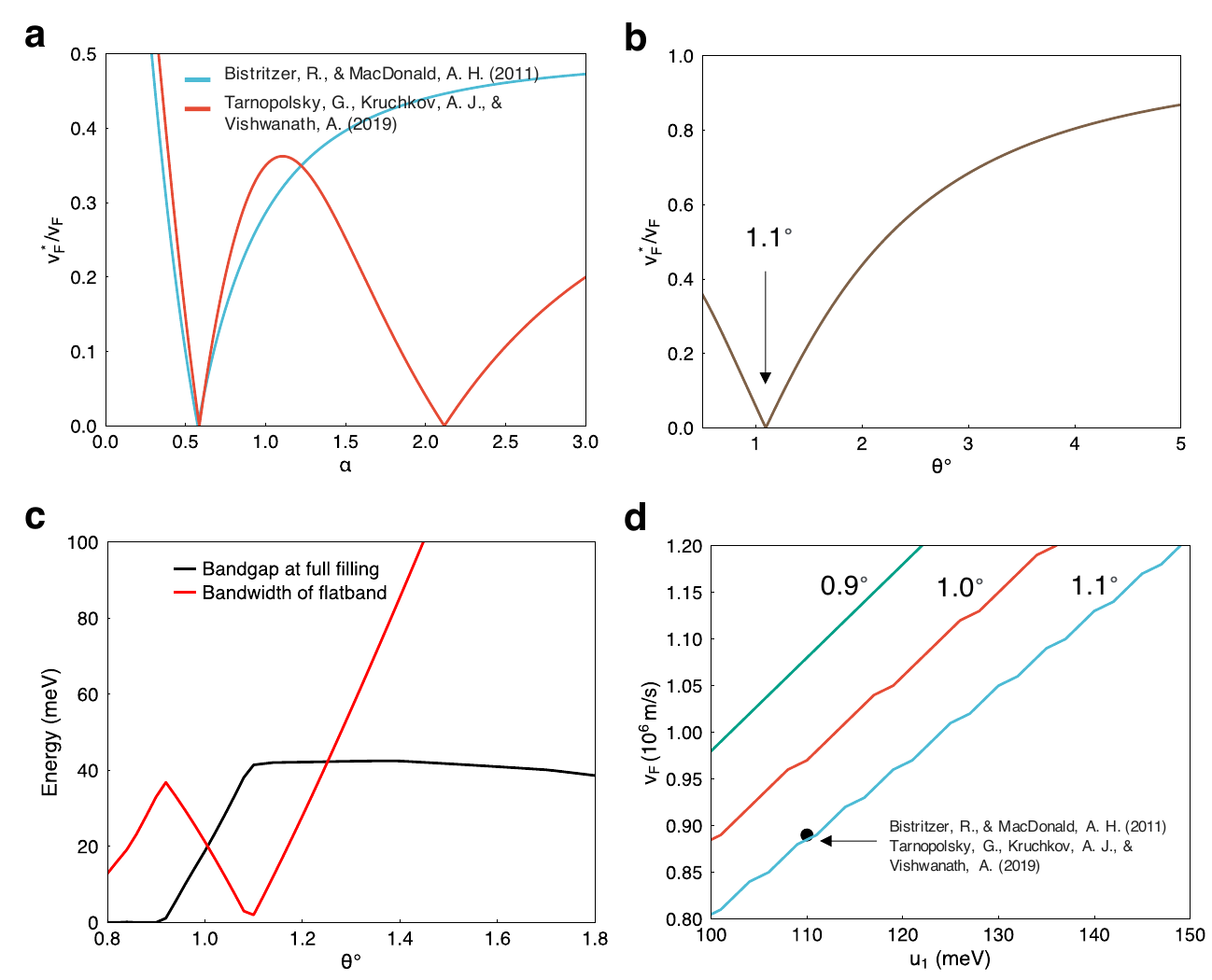}
    \caption{\textbf{Magic angle} \textbf{a,} The Fermi velocity of TBG within the flat band ($v_F^*$) normalized to $v_F$ versus the dimensionless parameter $\alpha$=$u_1$/$v_F$$k_\theta$ reproduced from ref\cite{bistritzer2011moire,tarnopolsky2019origin}, where $v_F$ is the Fermi velocity of monolayer graphene that defines the intralayer Hamiltonian, $k_\theta$ is the wave vector defined by the moiré superlattice, and $u_1$ is the interlayer coupling energy at AB site. The first magic angle is predicted to be at $\alpha$ $\approx$ 0.586, corresponding to magic angle 1.1\degree. \textbf{b,} $v_F^*$/$v_F$ versus the twist angle ($\theta$) by converting $\alpha$ in \textbf{a} to $\theta$. \textbf{c,} The bandgap between the flat and remote bands and the bandwidth of the flat band as a function of the twist angle, where the bandgap reaches a maximum and the bandwidth reaches a minimum close to the magic angle. \textbf{d,} The magic angle as a function of $u_1$ and $v_F$. Varying the ratio $u_1$/$v_F$ results in different magic angles theoretically. The black dot indicates parameters adapted from ref\cite{bistritzer2011moire,tarnopolsky2019origin}.}
    \label{fig:SI_Alan}  
\end{figure*}

Around the magic angle (1.1\degree), the moiré periodicity is much larger than the lattice constant of monolayer graphene, resulting in each moiré period containing thousands of carbon atoms. The continuum model provides a simplified description of this system, allowing the calculation of the band structure with a few essential parameters: the twist angle ($\theta$), the Fermi velocity ($v_F$) of monolayer graphene, and the interlayer coupling energies at AA/AB ($u_0$/$u_1$) sites. The dimensionless parameter $\alpha$ is defined as $\alpha$=$u_1$/$v_F$$k_\theta$\cite{bistritzer2011moire,tarnopolsky2019origin}, where $k_\theta$=(4$\pi$/3$a_0$)$\cdot$2sin($\theta$/2) is the wave vector defined by the moiré superlattice, and $a_0$ is the lattice constant of monolayer graphene. The magic angle is theoretically found when the Fermi velocity of the flat band ($v_F^*$) reaches minimum\cite{bistritzer2011moire,tarnopolsky2019origin}, where  $\alpha$$\approx$0.586 for the first magic angle (Fig. \ref{fig:SI_Alan}a). By converting $\alpha$ to $\theta$, the magic angle is 1.1\degree\ (Fig. \ref{fig:SI_Alan}b) when $v_F$=0.89$\times$10$^6$ m/s and  $u_1$=110 meV as adapted by previous work\cite{bistritzer2011moire,tarnopolsky2019origin}. In addition, the bandwidth of the flat band and the bandgap between the flat and remote bands are also indications of the magic angle \cite{carr2019exact}, where the bandwidth reaches a minimum and the bandgap reaches a maximum (Fig. \ref{fig:SI_Alan}c) .

Experimental works\cite{saito2020independent,cao2021nematicity} show the critical temperature of superconductivity and the correlated bandgap at $\nu$ = -2 versus twist angle reaches the maxima at around 1.1\degree, as an indication of the magic angle where the electron-electron interactions are the strongest. 
However, fixing the magic angle to 1.1\degree\ doesn't provide a unique mathematical solution for the choice of $v_F$ and $u_1$. The magic angle can be maintained by varying $v_F$ and $u_1$ while fixing the ratio $u_1$/$v_F$ as shown in Fig. \ref{fig:SI_Alan}d. For monolayer graphene, $v_F$ is directly related to the nearest hopping energy ($t$) between sub-lattices ($v_F$=$\sqrt{3}$$t$$a_0$/2\cite{neto2009electronic}).

\begin{figure*}[h]
    \centering
    \includegraphics[width=0.9\textwidth]{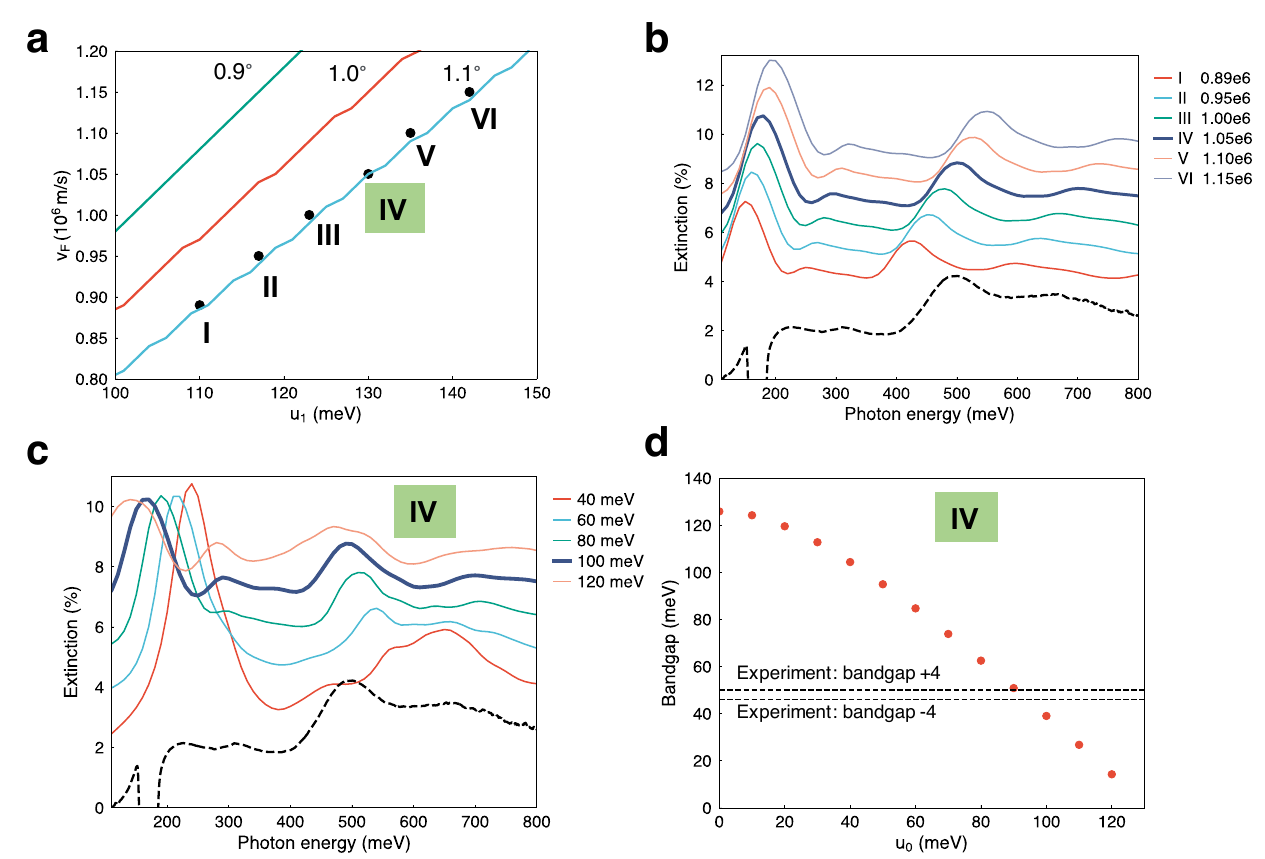}
    \caption{\textbf{Modelling the band structure of 1.46\degree.} \textbf{a,} Spots $I$ to $VI$ indicate that we vary $v_F$ and $u_1$ simultaneously while fixing the magic angle to be 1.1\degree (i.e. fixing the ratio $u_1$/$v_F$). \textbf{b,} The calculated extinction spectra for Spots $I$ to $VI$.  $u_0$/$u_1$ is fixed to be 70\%. The black dashed line is the experimental spectrum of 1.46\degree. Spot $IV$ ($v_F$=1.05$\times$$10^6$ m/s, $u_1$=130 meV) shows the best agreement with the experimental spectrum. \textbf{c,} The calculated extinction spectrum of 1.46\degree\ with $u_0$ from 40 meV to 120 meV, while $v_F$ and $u_1$ are fixed at Spot $IV$. $u_0$=100 meV gives the best agreement. \textbf{d,} The red dots represent the bandgap at $\nu$=$\pm$4 versus $u_0$. The black dashed lines indicate the bandgap extracted from transport measurements. $u_0$=100 meV gives a close match to the experiment value. }
    \label{fig:SI_Alan_1p46}  
\end{figure*}

We start with 1.46\degree. We note that we find the best solution by varying $u_0$ and $v_F$ simultaneously while fixing $u_1$/$v_F$ by maintaining the magic angle 1.1\degree (Fig. \ref{fig:SI_Alan_1p46}a). Fig. \ref{fig:SI_Alan_1p46}b shows the calculated extinction spectra for Spot $I$ to $VI$ as indicated in Fig. \ref{fig:SI_Alan_1p46}a, while keeping $u_0$/$u_1$=70\% and the magic angle to be 1.1\degree. By changing $v_F$, the shape of the optical spectrum remains the same but the energies of absorption peaks shift to higher energy with increasing $v_F$. Fig. \ref{fig:SI_Alan_1p46}c shows the calculated spectra by varying $u_0$ while keeping $u_1$ and $v_F$ at Spot $IV$. The shape of the spectrum is modified dramatically by varying $u_0$/$u_1$. Fig. \ref{fig:SI_Alan_1p46}d shows that the calculated bandgap between the flat and remote band decreases with increasing $u_0$/$u_1$. Using the calculations presented in Fig. \ref{fig:SI_Alan_1p46}, we find the best match between experiments and theory and extract the continuum model parameters in the following way. First, we fix the magic angle to be 1.1\degree\  by fixing the ratio $u_1$/$v_F$. We start with a tentative value $v_F$=1.05$\times$10$^6$ m/s. Second, we vary $u_0$/$u_1$ (Fig. \ref{fig:SI_Alan_1p46}c) and find the best agreement with the experimental spectrum in terms of the spectral shape, while the offset in the peak positions can be adjusted in the next step by varying $v_F$. In this step, we extract $u_0$/$u_1$. Third, we vary $v_F$ while keeping the magic angle 1.1\degree\  (Fig. \ref{fig:SI_Alan_1p46}a,b) and keeping the $u_0$/$u_1$ as extracted from the second step. In this step, we extract $v_F$ and $u_1$. Finally, After extracting $u_0$, $u_1$, and $v_F$ from the previous steps, we use the same parameters to calculate the bandgap between the flat and remote band, which is then compared with temperature-dependent transport measurements as a final check. We extract $u_0$=100 meV, $u_1$=130 meV, and $v_F$=1.05$\times$10$^6$ m/s for 1.46\degree. We note that we take priority of achieving the best agreement to the optical spectrum compared to Arrhenius extracted gaps. We try to reproduce the trend for the bandgap versus the twist angle, in which the bandgap reaches its maximal values at around 1.1\degree\ (Fig \ref{fig:Fig_4}a).

\begin{figure*}[!h]
    \centering
    \includegraphics[width=0.9\textwidth]{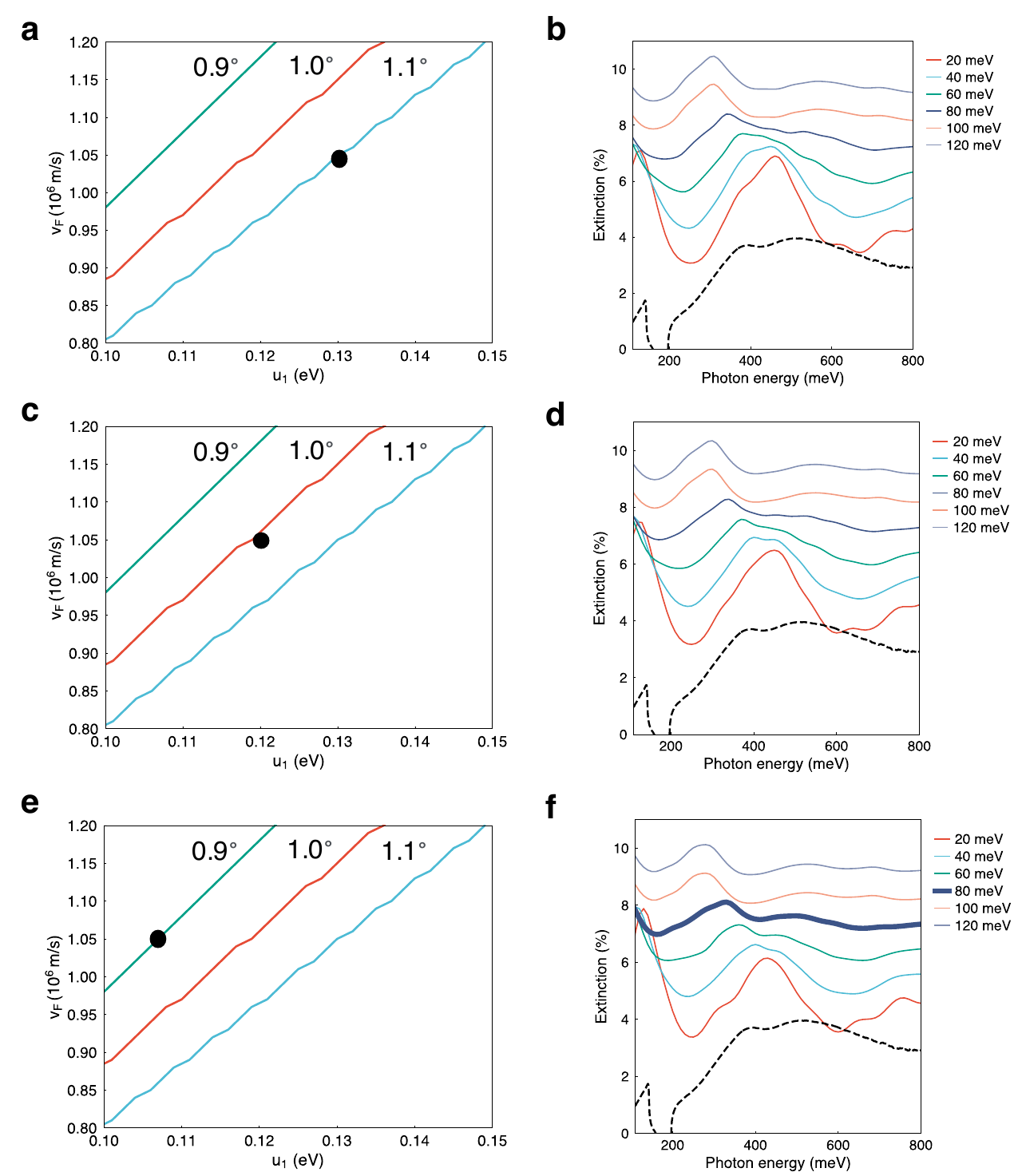}
    \caption{\textbf{Modelling 0.94\degree\  under different magic angles.} The black dots in \textbf{a}, \textbf{c}, and \textbf{e} indicate the choice of $u_1$ and $v_F$ for calculating the extinction spectrum as displayed in \textbf{b}, \textbf{d}, and \textbf{f}, respectively. We fix $v_F$=1.05$\times$10$^6$ m/s and vary $u_1$ to change the magic angle (1.1\degree, 1.0\degree, and 0.9\degree). \textbf{b}, \textbf{d}, and \textbf{f} show the calculated extinction spectra by varying $u_0$ while $u_1$ and $v_F$ are fixed as shown in \textbf{a}, \textbf{c}, and \textbf{e}. The black dashed lines are the experimental spectra of 0.94\degree. We find a good agreement in \textbf{f} for $u_0$ = 80 meV when the magic angle is 0.9\degree. At this step, we just consider the relative spectral shape, exact peak positions will be adjusted later by tunning $v_F$. }
    \label{fig:SI_Alan_0p94}  
\end{figure*}

\begin{figure*}[!h]
    \centering
    \includegraphics[width=1\textwidth]{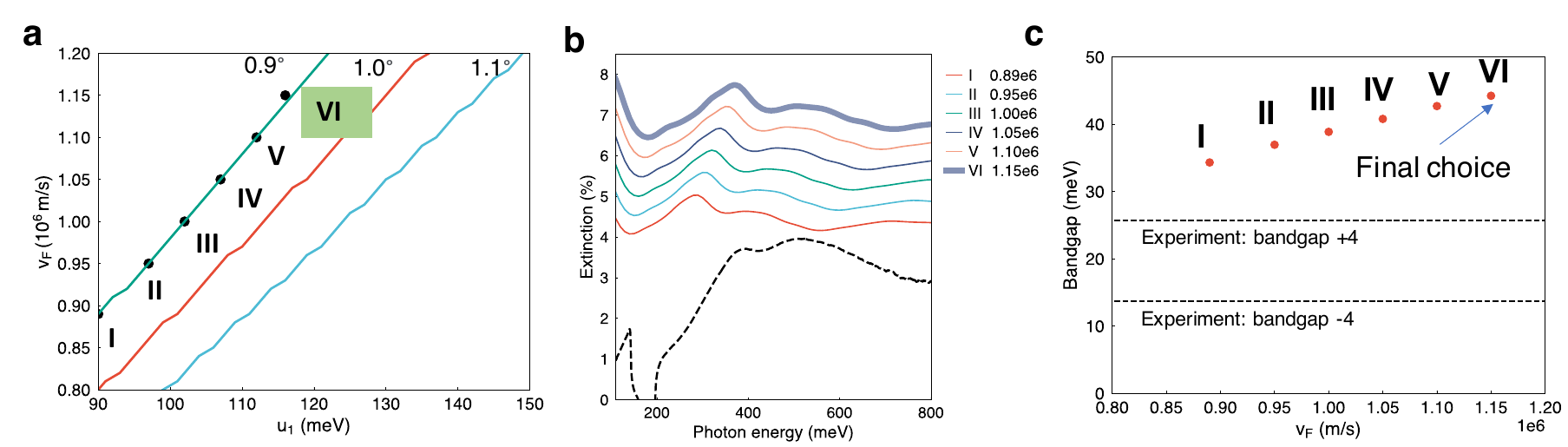}
    \caption{\textbf{Modelling 0.94\degree\  by changing $v_F$ at magic angle 0.9\degree.} \textbf{a,} Spots $I$ to $VI$ indicate varying $v_F$ and $u_1$ simultaneously while fixing the magic angle to be 0.9\degree. \textbf{b,} The calculated extinction spectra for Spots $I$ to $VI$ as shown in \textbf{b}, while keeping $u_0$/$u_1$ $\approx$ 70\% as extracted from Fig. \ref{fig:SI_Alan_0p94}. The black dashed line is the experimental spectrum of 0.94\degree. We find the best agreement when $v_F$=1.15$\times$$10^6$ m/s, $u_1$=115 meV (spot $VI$). \textbf{c,} The red dots show the bandgap at $\nu$=$\pm$4 calculated for Spots $I$ to $VI$. The black dashed lines indicate experimental values. }
    \label{fig:SI_Alan_0p94_vf}  
\end{figure*}

We can find good agreements between the experiments and the continuum model calculations for twist angles 1.12\degree-1.67\degree\  by following the above-mentioned procedures. We extract $u_1$=130 meV, and $v_F$=1.05$\times$10$^6$ m/s (corresponding to magic angle 1.1\degree) for 1.67\degree-1.12\degree. However, following the same procedure would lead to poor agreements for 0.94\degree\ and 0.98\degree (Fig. \ref{fig:SI_Alan_0p94}a,b). Ref\cite{carr2019exact, bennett2023twisted} suggests that the magic angle may not be a single value, but an angular range by taking into account the lattice relaxation. Inspired by this theoretical prediction, we shift the magic angle from 1.1\degree\ to 1.0\degree\ (Fig. \ref{fig:SI_Alan_0p94}c,d) and 0.9\degree\ (Fig. \ref{fig:SI_Alan_0p94}e,f). Here, we focus on 0.94\degree, and 0.98\degree\ follows the same procedure. When the magic angle is 1.1\degree\  and 1.0\degree, we cannot find agreement between the calculated and measured spectrum considering the spectral shape in Fig. \ref{fig:SI_Alan_0p94}b,d (as discussed previously, here we are at the second step). Interestingly, we find that when the magic angle is 0.9\degree\ and $u_0$=80 meV, we can achieve a good match between the calculated and measured spectrum just considering the spectral shape as shown in Fig. \ref{fig:SI_Alan_0p94}e,f (the offset in energy will be adjusted by the following steps). Fig. \ref{fig:SI_Alan_0p94_vf}a,b,c shows the procedure (the third step) to extract $u_1$ and $v_F$. We notice that in Fig. \ref{fig:SI_Alan_0p94_vf}c, there is around a 20 meV deviation between the experiment and calculation, which is acceptable considering good agreements with the experimental optical spectrum. 

To conclude, we achieve good agreement between the continuum model in a single-particle picture with only essential parameters and our experimental data. The continuum model parameters used throughout this work are displayed in Table \ref{Table:parameters}, the same as Fig. \ref{fig:Fig_4}b.

\subsection{Cryogenic temperature FTIR transmission measurement}
\label{sec:cryoFTIR}

\begin{figure*}[h]
    \centering
    \includegraphics[width=0.5\textwidth]{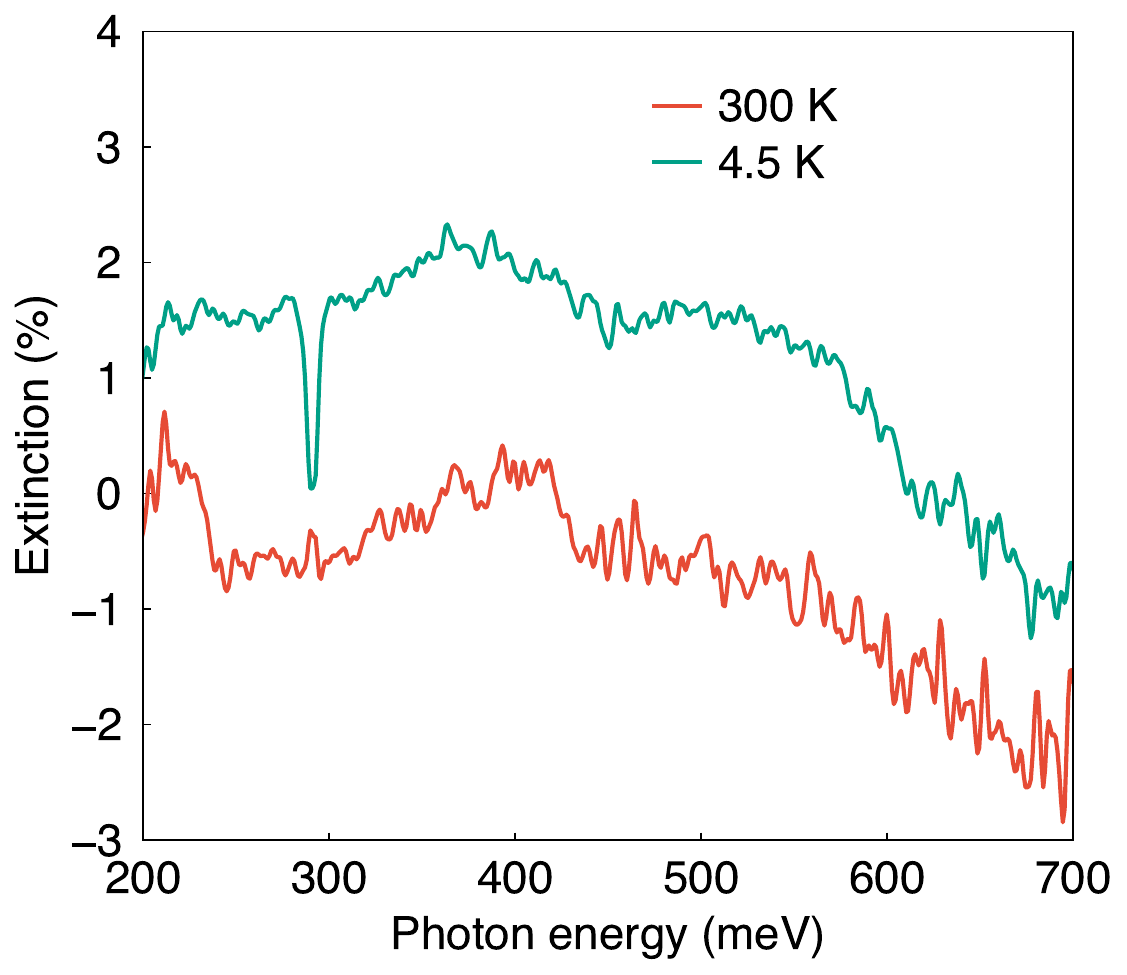}
    \caption{\textbf{Low temperature FTIR spectrum.}  Extinction spectra of a 1.16\degree\  device at cryogenic temperature (4.5\,K, green) and 300\,K (red). The sharp dip (291.3 meV) in the spectrum is due to the absorption of carbon dioxide in the air, which can be safely ignored. }
    \label{fig:SI_cryo_FTIR}  
\end{figure*}

In the main text, we employ the continuum model in a single-particle picture without including electron-electron interactions. To support this claim, we perform infrared FTIR measurements of a 1.16\degree\ stack at 4.5 K and 300 K  (Fig. \ref{fig:SI_cryo_FTIR}), considering that electron-electron interactions may become prominent at cryogenic temperature and therefore modify the optical transition energies significantly. However, no notable spectral shift is observed and the two absorption peaks (corresponding to transitions 3 and 4 discussed in the main text) stay at the same energy at both temperatures. This justifies our choice of a single-particle picture. We note that the cryogenic temperature FTIR measurement is performed at the setup described in ref\cite{nedoliuk2019colossal}.

\subsection{Extinction spectra peak extraction}
\label{sec:peak_extraction}

\begin{figure*}[!h]
    \centering
    \includegraphics[width=0.5\textwidth]{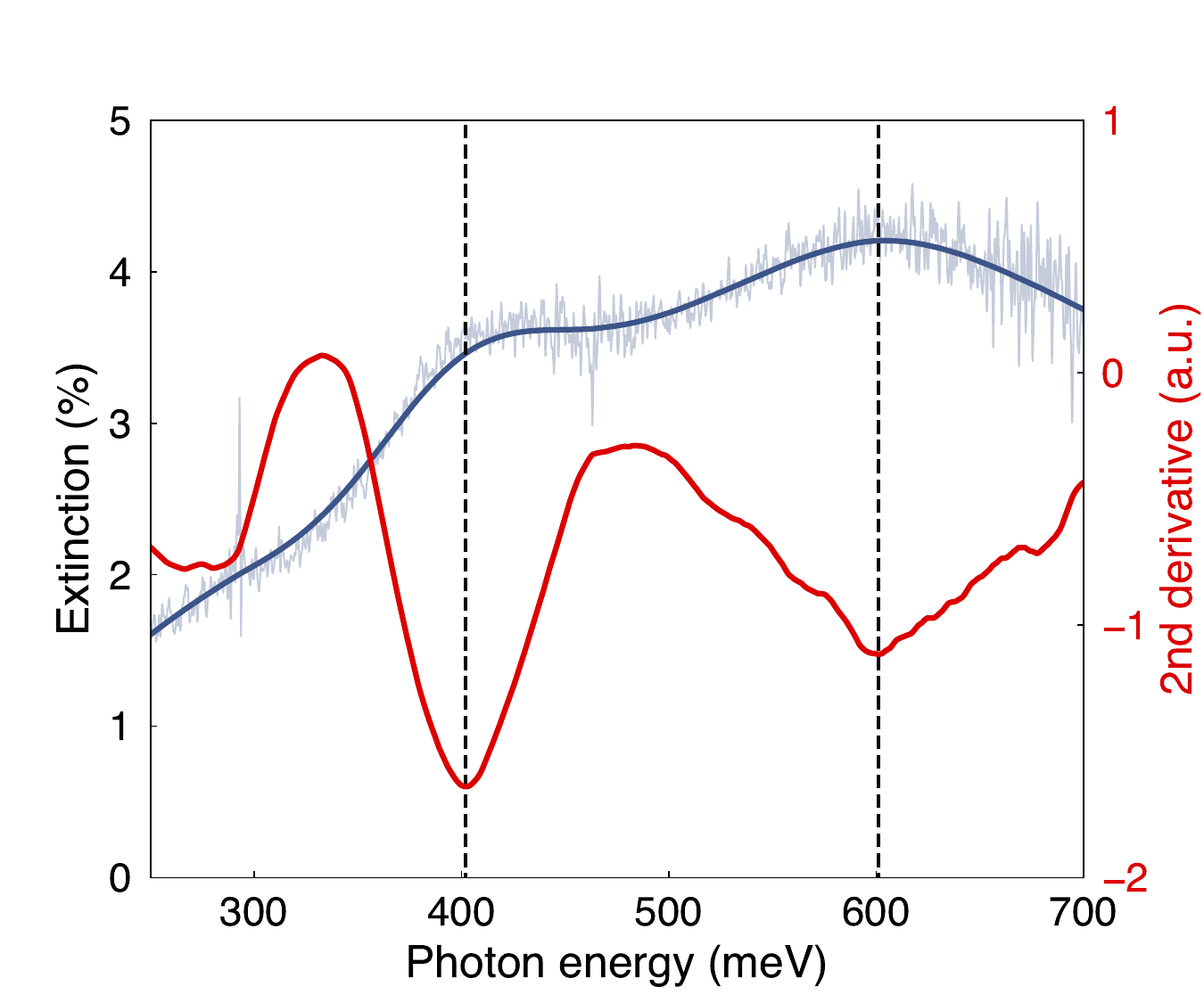}
    \caption{\textbf{Extraction of peak positions for 1.16\degree\ TBG.} We extract the positions of the absorption peaks in the extinction spectrum by finding the minima of its second derivative\cite{griffiths2007fourier}. Blue line: raw and smoothed spectrum. Solid red line: second derivative of the extinction. Black dashed lines: extracted peak positions from the second derivative minima.}
    \label{fig:SI_2nd derivative}  
\end{figure*}

The FTIR measurements presented in this work are all performed at room temperature unless otherwise stated. Under such conditions, most of the observed spectral peaks are broad, with exact peak positions hard to identify. The peak positions, such as the one shown in Fig. \ref{fig:Fig_3}d, are found from the minima of the second derivative\cite{griffiths2007fourier}. Fig. \ref{fig:SI_2nd derivative} shows an example for such procedure.


\begin{table*}[h!]
\caption{Alignment for hBN}\label{Table:alignment}%
\begin{tabular}{@{}llll@{}}
\toprule
Twist angle & top hBN/bottom hBN& top hBN/graphene & graphene/bottom hBN\\
\midrule
0.94$\degree$ & 9$\degree$ & 4$\degree$ &17$\degree$\\
0.98$\degree$ & 17$\degree$ & 4$\degree$ &13$\degree$\\
1.12$\degree$ & 8$\degree$ & 5$\degree$ &3$\degree$\\
1.16$\degree$ & 3$\degree$ & 7$\degree$ &10$\degree$\\
1.46$\degree$ & 1$\degree$ & 5$\degree$ &6$\degree$\\
1.67$\degree$ & 13$\degree$ & 10$\degree$ &23$\degree$\\
\botrule
\end{tabular}
\end{table*}

\begin{table*}[h]
\caption{The continuum model parameters (the same as Fig. \ref{fig:Fig_4}b)  we used throughout this work for the best agreement to FTIR and transport measurements. }\label{Table:parameters}
\begin{tabular}{@{}lllll@{}}
\toprule
Twist angle & $u_0$ (meV) & $u_1$ (meV) & $v_F$ (m/s)& Magic-angle\\
\midrule
0.94$\degree$ & 80 & 115 & 1.15$\times$10$^6$& 0.886\degree\\
0.98$\degree$ & 80 & 120 & 1.15$\times$10$^6$& 0.924\degree\\
1.12$\degree$ & 80 & 130 & 1.05$\times$10$^6$& 1.096\degree\\
1.16$\degree$ & 90 & 130 & 1.05$\times$10$^6$& 1.096\degree\\
1.46$\degree$ & 100 & 130 & 1.05$\times$10$^6$& 1.096\degree\\
1.67$\degree$ & 100 & 130 & 1.05$\times$10$^6$& 1.096\degree\\
\botrule
\end{tabular}
\end{table*}

 \section*{References}
\bibliography{FTIR_Geng}

\end{document}